\documentclass[aps,preprint,nofootinbib,floatfix,superscriptaddress]{revtex4}
\usepackage{graphicx,xcolor}
\usepackage[utf8]{inputenc}
\usepackage{amsmath,amssymb}
\usepackage[pdfusetitle]{hyperref}
\usepackage{epstopdf}
\usepackage{slashed}

\newcommand{\lsim}{\mathrel{\mathop{\kern 0pt \rlap
  {\raise.2ex\hbox{$<$}}}
  \lower.9ex\hbox{\kern-.190em $\sim$}}}
\newcommand{\gsim}{\mathrel{\mathop{\kern 0pt \rlap
  {\raise.2ex\hbox{$>$}}}
  \lower.9ex\hbox{\kern-.190em $\sim$}}}

\newcommand{\be}{\begin{equation}}
\newcommand{\ee}{\end{equation}}
\newcommand{\ba}{\begin{array}}
\newcommand{\ea}{\end{array}}
\newcommand{\bea}{\begin{eqnarray}}
\newcommand{\eea}{\end{eqnarray}}

\newcommand{\figscale}{0.4}

\def  \bcen   {\begin{center}}
\def  \ecen   {\end{center}}
\def  \beq    {\begin{equation}}
\def  \eeq    {\end{equation}}
\def  \nn     {\nonumber }

\def\la   {\lambda}

\def\nn{\nonumber}
\def\lee { \left( }
\def\rii { \right) }
\def\lan   {\langle}
\def\ran   {\rangle}

\def\to {\rightarrow}
\def\lphp {\la^\prime_{H\Phi}}

\begin{document}

\title{A Sub-GeV Low Mass Hidden Dark Sector of $SU(2)_H \times U(1)_X$}
\author{Raymundo Ramos}
\email{raramos@gate.sinica.edu.tw}
\affiliation{Institute of Physics, Academia Sinica, Nangang, Taipei 11529, Taiwan}
\author{Van Que Tran}
\email{vqtran@nju.edu.cn}
\affiliation{School of Physics, Nanjing University, Nanjing 210093, China}
\author{Tzu-Chiang Yuan}
\email{tcyuan@phys.sinica.edu.tw}
\affiliation{Institute of Physics, Academia Sinica, Nangang, Taipei 11529, Taiwan}

\date{\today}

\begin{abstract}
We present a detailed study of the non-abelian vector dark matter candidate $W^\prime$
with a MeV--GeV low mass range, 
accompanied by a dark photon $A^\prime$ and a dark $Z^\prime$ of similar masses,
in the context of a gauged two-Higgs-doublet model with the hidden gauge group that has the same structure 
as the Standard Model electroweak gauge group.
The stability of dark matter is protected by an accidental discrete $Z_2$ symmetry ($h$-parity) which was usually 
imposed {\it ad hoc} by hand.
We examine the model by taking into account various experimental constraints including dark photon searches
at NA48, NA64, E141, $\nu$-CAL, BaBar and LHCb experiments, 
electroweak precision data from LEP, 
relic density from Planck satellite, direct (indirect) detection of dark matter 
from CRESST-III, DarkSide-50, XENON1T (Fermi-LAT), and collider physics from the LHC. 
The theoretical requirements of bounded from below of the scalar potential and tree level perturbative 
unitarity of the scalar sector are also imposed.
The viable parameter space of the model consistent with all the constraints is exhibited.
While a dark $Z^\prime$ can be the dominant contribution in the relic density 
due to resonant annihilation of dark matter, a dark photon is crucial to dark matter direct detection.
We also demonstrate that the parameter space can be further probed by various sub-GeV direct dark matter 
experimental searches at CDEX, NEWS-G and SuperCDMS in the near future.
\end{abstract}

\maketitle

\section{Introduction\label{section:1}}

The stability of heavy dark matter (DM) is usually implemented in many particle physics models beyond the standard model (SM) 
by imposing a discrete $Z_2$ symmetry in the Lagrangian. These models include the simplest 
scalar phantom model by adding just a singlet $Z_2$-odd scalar field (real or complex) to the SM~\cite{Silveira:1985rk}, 
the popular inert two-Higgs-doublet model (I2HDM)~\cite{Deshpande:1977rw,Ma:2006km}, 
the minimal supergravity standard model with 
$R$-parity (MSSM)~\cite{Chamseddine:1982jx,Nath:1982zq,Nilles:1983ge}, 
little Higgs model with $T$-parity (LHM)~\cite{Cheng:2003ju,Cheng:2004yc,Low:2004xc}  {\it etc}. 
In I2HDM, the DM candidate can be either the CP-even or -odd scalar
residing in the second $Z_2$-odd Higgs doublet. Many detailed analysis of DM phenomenology in the scalar phantom models and 
I2HDM can be found in the literature in~\cite{McDonald:1993ex,Burgess:2000yq,Cheung:2012xb} 
and~\cite{,Barbieri:2006dq,LopezHonorez:2006gr,Arhrib:2013ela,Belyaev:2016lok,Lu:2019lok,Fabian:2020hny} respectively.
In MSSM, the lightest supersymmetric particle (LSP)
for the DM candidate can be the spin 0 sneutrino (the superpartner of neutrino) or the lightest spin 1/2 neutralino~\cite{Akula:2011aa}
(in general a linear combination of two gauginos and two Higgsinos~\cite{Arnowitt:1995vg}). We note that 
in some low energy supergravity models, the LSP can be the spin 3/2 gravitino, the superpartner of graviton.
For a review of supersymmetric dark matter, see for example~\cite{Jungman:1995df}.
In LHM, the spin 1 $T$-odd partner of the photon can be the DM candidate whose collider implication was 
studied in~\cite{Chen:2006ie}. 
There are also well-motivated non-abelian dark matter models based on additional gauge group 
like $SU(2)$~\cite{Carone:2013wla,Davoudiasl:2013jma,Barman:2017yzr,Barman:2019lvm,Abe:2020mph,Hu:2021pln}, 
in which the extra spin 1 gauge boson $W^\prime$ can be a DM candidate. 
Moreover, instead of specifying an underlying dark matter model, one can also use the effective dark matter theory approach~\cite{Cao:2009uw,Goodman:2010ku,Cheung:2012gi} to discuss various dark matter phenomenologies~\cite{Goodman:2010yf,Goodman:2010qn,Cheung:2010ua,Cheung:2011nt,Huang:2019ikw,NathMaity:2021cne}.

Recently a gauged two-Higgs-doublet model (G2HDM)
based on an extended electroweak gauge group $SU(2)_L \times U(1)_Y \times SU(2)_H \times U(1)_X$,
was proposed~\cite{Huang:2015wts}, in which a hidden discrete $Z_2$ symmetry
($h$-parity)~\cite{Chen:2019pnt} arises naturally as an accidental 
symmetry rather than imposed by hand. 
This discrete symmetry ensures the stability of the DM candidate in G2HDM, which can be either a complex scalar 
(which in general is a linear combination of various fields in the model)
or a heavy neutrino $\nu^H$ or an extra gauge boson $W^{\prime (p,m)}$, 
all of which have odd $h$-parity.
We note that, unlike the left-right symmetric model~\cite{Mohapatra:1979ia,Mohapatra:1980yp}, 
the $W^{\prime (p,m)}$ in G2HDM do not carry electric charge.

The novel idea of G2HDM, as compared with many variants of general 2HDM~\cite{Branco:2011iw}, 
is that the two Higgs doublets $H_1$ and $H_2$ of $SU(2)_L$ are 
grouped into a fundamental representation of a new gauge group $SU(2)_H$. 
Consistency checks of the model were scrutinized for the scalar and gauge sectors in~\cite{Arhrib:2018sbz}
and~\cite{Huang:2019obt} respectively.
In~\cite{Chen:2019pnt}, a detailed phenomenological analysis of the scalar DM candidate in G2HDM was carried out.
In general, the scalar DM candidate is a complex field made up of a linear combination of the components 
from the inert Higgs doublet $H_2$ and the $SU(2)_H$ doublet $\Phi_H$ and triplet $\Delta_H$. 
By performing a detailed parameter scan in the model it was demonstrated~\cite{Chen:2019pnt} that only the triplet-like DM 
is favored when all the constraints from the relic density, direct and indirect searches for the DM are taken into account. 
As discussed in~\cite{Huang:2015wts}, the triplet $\Delta_H$ plays the primary role as a trigger for 
spontaneous symmetry breaking of the model down to $U(1)_{\rm EM}$. 
We note also that this triplet can have topological implications from the hidden sector. 
Since $\Delta_H$ is an adjoint representation of $SU(2)_H$, 
there exists magnetic monopole~\cite{tHooft:1974kcl,Polyakov:1974ek} and dyon~\cite{Julia:1975ff} solutions 
in the hidden sector which may play the role of topological stable DM~\cite{Baek:2013dwa}.
However, the triplet is not required to generate realistic mass spectra for
all the particles in G2HDM\@. 
Therefore if one omits this triplet scalar, the scalar DM candidate is no longer 
favorable in the parameter space of G2HDM according to the analysis in~\cite{Chen:2019pnt}. 
However, as mentioned above, there are two other alternative DM candidates in the model.
In this paper, we will show that the non-abelian gauge boson $W^{\prime (p,m)}$ associated with  
$SU(2)_H$ can be a viable DM as well. In particular we will focus on the
low mass DM scenario in the MeV--GeV range
which has attracted a lot of attention in recent years. 

This paper is organized as follows.
In Sec.~\ref{sec:model}, we will review some salient features of the simplified  
G2HDM without introducing the Higgs triplet field $\Delta_H$ of the extra $SU(2)_H$. 
The theoretical constraints on the Higgs potential, electroweak precision data for the $Z$-boson mass shift from 
Large Electron-Positron Collider (LEP) data,
dark photon constraints from various low energy experiments 
and the 125 GeV Higgs data constraints from the Large Hadron Collider (LHC)
are discussed in Sec.~\ref{sec:constraints}\@.
We then turn to the experimental constraints of dark matter physics in
Sec.~\ref{sec:DMPheno}\@.
The cosmological relic density from Planck satellite~\cite{Aghanim:2018eyx}, 
underground direct detection constraints from 
CRESST III~\cite{Angloher:2017sxg},
DarkSide-50~\cite{Agnes:2018ves} and XENON1T~\cite{Aprile:2019xxb}, 
astrophysical gamma-ray indirect detection constraints from 
Fermi-LAT~\cite{Ackermann:2015zua, Fermi-LAT:2016uux}
and mono-jet constraints from LHC~\cite{Aaboud:2017phn, ATLAS:2020wzf,Sirunyan:2017hci}
for the sub-GeV dark matter $W^{\prime (p,m)}$ in G2HDM
are studied in Secs.~\ref{sec:relic},~\ref{sec:DD},~\ref{sec:ID} and~\ref{sec:monojet} respectively. 
Our numerical results are presented in Sec.~\ref{sec:results}\@. 
We conclude in Sec.~\ref{sec:conclusion}\@. In Appendix~\ref{appendixA}, we discuss the mixing effects in the 
gauge fixings of the model in a general renormalizable $R_\xi$ gauge.
This work can be regarded as an expanded detailed version of the compact and partial results presented in~\cite{Ramos:2021omo}.

\section{The Simplified G2HDM Model\label{sec:model}}

In this section, we discuss a simplified version of 
the G2HDM first proposed in~\cite{Huang:2015wts}.
In particular, we will remove the triplet scalar $\Delta_H$ in the original model
because it is not absolutely required for a realistic particle spectra
and the number of free parameters in the scalar potential can be reduced significantly.
We note that the Yukawa couplings are not affected by this simplification
since the triplet does not couple to the fermions in the model. 

\subsection{Particle Content\label{sec:particle_content}}

The gauge group of the simplified G2HDM is the same as in~\cite{Huang:2015wts},
\begin{equation}
{\mathcal G} =  SU(3)_{C}\times SU(2)_{L} \times U(1)_{Y}  \times SU(2)_{H} \times U(1)_X \; . \nonumber
\end{equation}

In Table~\ref{tab:quantumnos}, we summarize the matter content and their quantum number assignments in G2HDM\@.
At the minimum risk of confusion, we will continue refer this model as G2HDM to avoid cluttering throughout 
the paper with the adjective word ``simplified''. What we are really
concerned about is the electroweak part of $\mathcal G$, 
so the color group $SU(3)_C$ is not relevant in what follows.


%
\begin{table}[htbp!]
\centering
\begin{tabular}{|c|c|c|c|c|c|c|}
\hline
Matter Fields & $SU(3)_C$ & $SU(2)_L$ & $SU(2)_H$ & $U(1)_Y$ & $U(1)_X$ & $h$-parity \\
\hline \hline
$Q_L=\left( u_L \;\; d_L \right)^{\rm T}$ & 3 & 2 & 1 & 1/6 &0 & $+ \; +$ \\
$U_R=\left( u_R \;\; u^H_R \right)^{\rm T}$ & 3 & 1 & 2 & 2/3 & 1 & $+ \; -$ \\
$D_R=\left( d^H_R \;\; d_R \right)^{\rm T}$ & 3 & 1 & 2 & $-1/3$ &$ -1$  & $- \; +$ \\
\hline
$u_L^H$ & 3 & 1 & 1 & 2/3 & 0 & $-$ \\
$d_L^H$ & 3 & 1 & 1 & $-1/3$ & 0 & $-$ \\
\hline
$L_L=\left( \nu_L \;\; e_L \right)^{\rm T}$ & 1 & 2 & 1 & $-1/2$ & 0 & $+ \; +$ \\
$N_R=\left( \nu_R \;\; \nu^H_R \right)^{\rm T}$ & 1 & 1 & 2 & 0 & 1 & $+ \; -$ \\
$E_R=\left( e^H_R \;\; e_R \right)^{\rm T}$ & 1 & 1 & 2 &  $-1$  & $-1$ & $- \; +$ \\
\hline
$\nu_L^H$ & 1 & 1 & 1 & 0 & 0 & $-$\\
$e_L^H$ & 1 & 1 & 1 & $-1$  & 0 & $-$\\
\hline\hline
$H=\left( H_1 \;\; H_2 \right)^{\rm T}$ & 1 & 2 & 2 & 1/2 & 1 & $+ \; -$\\
$\Phi_H=\left( \Phi_1 \;\; \Phi_2 \right)^{\rm T}$ & 1 & 1 & 2 & 0 & 1 & $- \; +$\\
$\mathcal S$ & 1 & 1 & 1 & 0 & 0 & $+$\\
\hline
\end{tabular}
\caption{\label{tab:quantumnos}Matter content and their quantum number assignments in G2HDM\@. 
The electric charge $Q$ in unit of positron charge $e$ is given by $Q=T^3_L + Y$. The scalar $\mathcal S$ in the last row
is the Stueckelberg field introduced in~\cite{Huang:2015wts} to give mass for the $U(1)_X$ gauge boson.
}
\end{table}

\subsection{Higgs Potential and Spontaneous Symmetry Breaking}

The most general Higgs potential which is invariant under $SU(2)_L\times U(1)_Y \times SU(2)_H \times U(1)_X$
can be written down as follows
\begin{align}\label{eq:V}
V = {}& - \mu^2_H   \left(H^{\alpha i}  H_{\alpha i} \right)
+  \lambda_H \left(H^{\alpha i}  H_{\alpha i} \right)^2  
+ \frac{1}{2} \lambda'_H \epsilon_{\alpha \beta} \epsilon^{\gamma \delta}
\left(H^{ \alpha i}  H_{\gamma  i} \right)  \left(H^{ \beta j}  H_{\delta j} \right)  \nn \\
{}&- \mu^2_{\Phi}   \Phi_H^\dag \Phi_H  + \la_\Phi \lee \Phi_H^\dag \Phi_H  \rii^2 
+\lambda_{H\Phi} \lee H^\dag H  \rii  \lee \Phi_H^\dag \Phi_H \rii  
 + \lambda^\prime_{H\Phi} \lee H^\dag \Phi_H  \rii  \lee \Phi_H^\dag H \rii, 
\end{align}
where  ($\alpha$, $\beta$, $\gamma$, $\delta$) and ($i$, $j$) refer to the $SU(2)_H$ and $SU(2)_L$ indices respectively, 
all of which run from 1 to 2, and $H^{\alpha i} = H^*_{\alpha i}$.

To facilitate spontaneous symmetry breaking (SSB) and obtain the particle mass spectra of the model,
we shift the fields based on our conventional wisdom
\begin{eqnarray}
\label{eq:scalarfields}
H_1 \equiv 
\begin{pmatrix}
H_{11} \\ H_{12}
\end{pmatrix}
=
\begin{pmatrix}
G^+ \\ \frac{v + h}{\sqrt 2} + i \frac{G^0}{\sqrt 2}
\end{pmatrix}
, \;
H_2 \equiv 
\begin{pmatrix}
H_{21} \\ H_{22}
\end{pmatrix}
= 
\begin{pmatrix}
H^+ \\ H_2^0
\end{pmatrix}
, \;
\Phi_H = 
\begin{pmatrix}
G_H^p \\ \frac{v_\Phi + \phi_2}{\sqrt 2} + i \frac{G_H^0}{\sqrt 2}
\end{pmatrix}
, \;
\end{eqnarray}
where $v$ and $v_\Phi$ are the vacuum expectation values (VEV) of $H_1$ and $\Phi_{H}$ fields respectively.
$H_2$ is the inert doublet in G2HDM and hence does not have VEV.
Naively we would think that the Goldstone bosons $G^+$, $G^p_H$, $G^0$ 
and $G^0_H$ will be absorbed by the longitudinal components 
of $W^+$, $W^{\prime p}$, $W^3$ and $W^{\prime 3}$ respectively. 
In Appendix~\ref{appendixA} we will show that the last three Goldstone fields have 
mixing effects with other fields in the scalar sector that makes the situation more 
interesting but a little bit more complicated.

Substituting the scalar field decomposition of Eq.~(\ref{eq:scalarfields})
into the scalar potential Eq.~(\ref{eq:V}) and then minimize the potential, 
one can obtain the solutions of VEVs as follows
\begin{eqnarray}
\label{vevv}
v^2 & = & \frac{2 \left( \lambda_{H\Phi} \mu_\Phi^2 - 2 \lambda_{\Phi}  \mu_H^2 \right)}
{  \lambda_{H\Phi}^2 - 4 \lambda_H\lambda_\Phi }
 \; , \\
\label{vevphi}
v_\Phi^2 & = & \frac{2\left(\lambda_{H\Phi} \mu_H^2 - 2 \lambda_{H} \mu_\Phi^2\right)}
{  \lambda_{H\Phi}^2 - 4 \lambda_H\lambda_\Phi  }\; .
\end{eqnarray}

Equivalently, we can use the minimization conditions to trade $\mu_H^2$ and $\mu_\Phi^2$ with 
$v$ and $v_\Phi$ as
\begin{align}
\label{eq:muh2min}
\mu_H^2 & = \lambda_{H} v^{2} + \frac{\lambda_{H \Phi} v_{\Phi}^{2}}{2} \; , \\
\label{eq:muphi2min}
\mu_\Phi^2 & = \lambda_{\Phi} v_{\Phi}^{2} + \frac{\lambda_{H \Phi} v^{2}}{2} \; .
\end{align}


\subsection{Scalar Mass Spectrum\label{sec:scalar_mass}}

In the $S=\{h,\phi_2\}$ basis the mass matrix is given as 
\begin{equation}
{\mathcal M}_S^2 =
\begin{pmatrix}
	2 \lambda_H v^2 & \lambda_{H\Phi} v v_\Phi \\
	\lambda_{H\Phi} v v_\Phi & 2 \lambda_\Phi v_\Phi^2 
\end{pmatrix} \, .
\label{eq:scalarbosonmassmatrix}
\end{equation}
One can use an orthogonal transformation $O^S$, which can be parametrized as 
\begin{equation}
O^S = 
\begin{pmatrix}
	\cos \theta_1 & \sin \theta_1\\
	- \sin \theta_1 & \cos \theta_1 
\end{pmatrix} \,,
\label{ma:OS}
\end{equation}
where
\beq
\tan 2 \theta_1 = \frac{2\mathcal{M}^2_{S12}}{\mathcal {M}^2_{S22} - \mathcal{M}^2_{S11}} = \frac{\lambda_{H\Phi} v v_\Phi}{ \lambda_\Phi v_\Phi^2 - \lambda_H v^2 } \;,
\eeq
to diagonalize $\mathcal{M}_S^2$,
\beq\label{eq:mixH1}
\left(O^S\right)^{\rm T} \cdot {\mathcal M}_S^2 \cdot O^S 
= {\rm Diag}\left(m^2_{h_1}, m^2_{h_2}\right)\;,
\eeq
with $h_1$ being identified as the 125~GeV SM-like Higgs boson and $h_2$ as a
heavier scalar boson.
The mass squared eigenvalues of Eq.~\eqref{eq:scalarbosonmassmatrix} are
given by
\begin{equation}
m_{h_{1,2}}^2 = \lambda_{H} v^{2} + \lambda_{\Phi} v_{\Phi}^{2} \mp \sqrt{\lambda_{H}^{2}
		v^{4} - 2 \lambda_{H} \lambda_{\Phi} v^{2} v_{\Phi}^{2} + \lambda_{H
\Phi}^{2} v^{2} v_{\Phi}^{2} + \lambda_{\Phi}^{2} v_{\Phi}^{4}} \; .
\end{equation}

In the basis of $S'=\{G_H^p, H_2^{0*}\}$, we obtain the mass matrix:
\begin{equation}
{\mathcal M}_{S'}^{ 2} = \frac{1}{2}\lambda^\prime_{H\Phi}
\begin{pmatrix}
v^2 & v v_\Phi \\
v v_\Phi &  v_\Phi^2
\end{pmatrix} \; .
\label{eq:goldstonemassmatrix}
\end{equation}
Similarly, this mass matrix can be diagonalized by an orthogonal matrix, 
\begin{equation}
 O^{S'} = 
 \begin{pmatrix}
 \cos \theta_2 & \sin \theta_2\\
 - \sin \theta_2 & \cos \theta_2 
 \end{pmatrix} \,,
 \label{ma:OSp}
\end{equation}
where
\beq
\tan 2 \theta_2 = \frac{2\mathcal{M}^2_{S'12}}{\mathcal{M}^2_{S'22} - \mathcal{M}^2_{S'11}} = \frac{ 2 v v_\Phi}{ v_\Phi^2 - v^2 } \;,
\eeq
which gives
\beq\label{eq:mixH2}
\left(O^{S'}\right)^{\rm T} \cdot {\mathcal M}_{S'}^2 \cdot O^{S'}
= {\rm Diag}\left(0, m^2_{D}\right)\;.
\eeq
We note that, in Eq.~(\ref{eq:mixH2}), the zero eigenvalue corresponds to 
the Nambu-Goldstone boson mass eigenstate $\tilde G^p_H$, 
while the other eigenvalue
\begin{equation}
\label{eq:Dscalarmass}
m^2_{D}=\frac{1}{2}\lambda^\prime_{H\Phi} (v^2 + v_\Phi^2) \; ,
\end{equation}
is the mass of a new dark scalar boson. 

The charged Higgs boson mass is given as 
\beq\label{eq:chargedHiggsmass}
m^2_{H^\pm} =  \frac{1}{2}\left(\lambda^\prime_{H\Phi}v_\Phi^2 - \la^\prime_H v^2\right) \;.
\eeq
The Goldstone bosons $G^0$ and $G^0_H$ are massless.

The above scalar mass spectrum is derived in the so-called  't Hooft-Landau  gauge. 
We will discuss further the mixing effects of the Goldstone bosons 
with other scalar fields in a general renormalizable $R_\xi$ gauge in Appendix~\ref{appendixA}.

We note that $h_{1,2}$, $G^0$ and $G^0_H$ are even under $h$-parity, 
while $\tilde G^p_H$, $D$ and $H^\pm$ are odd~\cite{Chen:2019pnt}.
  
\subsection{Gauge Sector\label{sec:gauge_mass}}

After SSB, the $W^\pm$ gauge boson of $SU(2)_L$ remains the same as in SM with its mass given by
$m_W = g v/2$. 
The $SU(2)_H$ gauge boson $W^{\prime (p,m)}$ receives mass from $\lan H_1 \ran$
and $\lan \Phi_2 \ran$ given by
\begin{equation}
\label{eq:Wprimepmmass}
m_{W'} = \frac{1}{2} g_H \sqrt{v^2 + v^2_\Phi}\;.
\end{equation}
Note that $W^{\prime (p,m)}$ are electrically neutral
and thus do not mix with the SM $W^\pm$.
In addition, $W^{\prime (p,m)}$ is odd under $h$-parity. If it is the lightest $h$-parity odd particle in the model, 
it will be stable and can be a DM candidate. 

On the other hand, the SM neutral gauge bosons $B$ and $W^3$ can mix with the 
new gauge bosons $W^{\prime 3}$ and $X$, all of which have even $h$-parity. Together with the Stueckelberg mass parameters $M_X$ and $M_Y$
for the two abelian groups $U(1)_X$ and $U(1)_Y$, SSB generates a $4\times4$ neutral gauge boson 
mass matrix in the basis of $\left\{ B, W^3, W^{\prime 3}, X\right\}~$\cite{Huang:2015wts, Huang:2019obt}. 
Due to the theoretical motivations or prejudices mentioned in Ref.~\cite{Huang:2019obt}, 
we set the Stueckelberg mass $M_Y = 0$. 
Applying the weak rotation on upper left $2 \times 2$ block of the $4\times4$ mass matrix, 
one obtains immediately a zero eigenvalue identified as the SM photon 
and a $3\times3$ sub-matrix in the basis of $\left\{ Z^{\rm SM}, W^{\prime 3}, X \right\}$ 
given by~\cite{Huang:2019obt},
\be\label{eq:mZ33}
{\cal M}^2_Z = 
\begin{pmatrix} 
 m_{Z^\text{SM}}^2 &
        - \frac{g_H v }{2} m_{Z^\text{SM}} &
        - g_X v m_{Z^\text{SM}} \\
- \frac{g_H v}{2} m_{Z^\text{SM}} &
m_{W'}^2         &
        \frac{g_X g_H \left(v^{2} - v_\Phi^{2}\right)}{2}\\
 - g_X v m_{Z^\text{SM}} &
        \frac{g_X g_H \left(v^{2} - v_\Phi^{2}\right)}{2} &
        g_X^{2} (v^{2} + v_\Phi^{2}) + M_X^{2}
\end{pmatrix} \,,
\ee
where $g$, $g'$, $g_H$ and $g_X$ are the gauge couplings of
$SU(2)_L$, $U(1)_Y$, $SU(2)_H$ and $U(1)_X$ respectively, 
and $m_{Z^{\rm SM}} = v \sqrt{g^2 + g^{\prime 2}}/2$ is the SM $Z$ boson mass expression.
The mass matrix in Eq.~(\ref{eq:mZ33}) 
can be diagonalized by an orthogonal rotation matrix ${\cal O}$ 
so that
\footnote{The analytical expression of this rotation matrix can be found in
Ref.~\cite{Huang:2019obt}.}
\be\label{Oij}
\begin{pmatrix}
Z^{\rm SM} \\
W^{\prime 3}\\ 
X
\end{pmatrix} 
= {\cal O} \cdot 
\begin{pmatrix}
Z \\ Z' \\ A' 
\end{pmatrix}
 \;.
\ee
In this analysis, we arrange the neutral gauge boson masses as $m_{A'} < m_{Z'} < m_Z \simeq m_{Z^{\rm SM}}$
with $Z$ identified as the physical $Z$ boson with mass $91.1876\pm0.0021$ GeV~\cite{Zyla:2020zbs}---the
first heavy neutral vector gauge boson discovered in 1983 at the Super Proton Synchrotron at CERN\@.
Two more massive neutral vector gauge bosons are predicted in G2HDM\@. 

By means of a few assumptions motivated by our expectations, it is possible to
draw a few conclusions from Eq.~\eqref{eq:mZ33}.
First, the new gauge couplings $g_H$ and $g_X$ are expected to be much smaller
than the SM $g$ and $g'$ to avoid large effects on the very precise
measurements of the $Z$ properties.
Second, the scale of $v_\Phi$ is expected to be larger than 
$v$ given that it characterizes the scale of new physics and is directly
related to the masses of beyond the SM (BSM) states.
By neglecting any term composed by a product of any three or more of $g_H$,
$g_X$ and $v^2/v_\Phi^2$ it is possible to put Eq.~\eqref{eq:mZ33} into a
block diagonal matrix where only $W^{\prime 3}$ and $X$ mix resulting in the
approximation
\begin{equation}
\label{eq:mzApprox}
m_Z \approx m_{Z^{\rm SM}} \; .
\end{equation}
Moreover, the 2$\times$2 squared mass matrix of $W^{\prime 3}$ and $X$ can be
easily diagonalized and somewhat simple approximations can be found.
Since we want the hierarchy $m_{A'} < m_{Z'} < m_{Z^{\rm SM}}$, the $M_X$
parameter is required to have a value smaller than $v$.
Assuming $M_X < v$ allows us to expand the squared root in the general
solution for the eigenvalues of a 2$\times$2 matrix resulting in the
following approximations
\begin{align}
\label{eq:mzpApprox}
m_{Z'}^2 & \approx 
	m_{W'}^2 \left(1 + \frac{4 g_X^{2}}{g_H^2}\right)
	+ M_X^{2} \left[
		1 - \left(1 + \frac{4 g_X^{2}}{g_H^2} + \frac{M_X^{2}}{m_{W'}^2}\right)^{-1}
	\right] \; ,
\\
\label{eq:mapApprox}
m_{A'}^2 & \approx
M_X^2 \left(1 + \frac{4 g_X^{2}}{g_H^2} + \frac{M_X^{2}}{m_{W'}^2}\right)^{-1} \; ,
\end{align}
where $m_{W'}^2 \approx g_H^2 v_\Phi^2/4$ was used.
From these expressions we can see that $m_{Z'} \gtrsim m_{W'}$ and $m_{A'}\lesssim M_X$.

Since the couplings of the extra gauge bosons, $Z'$ and $A'$, to the SM fermions are proportional to 
the new gauge couplings $g_H$ and/or $g_X$ which are in general much smaller than the SM couplings 
$g$ and $g^\prime$, the Drell-Yan type processes are suppressed and 
this can explain the null results of BSM neutral gauge bosons searches at LEP\@.

In our study, the couplings of the extra gauge bosons to the SM charged
leptons and to quarks $u$ and $d$ will be important for dark photon
constraints and direct detection of DM\@.
The vectorial and axial parts of their couplings are given by
\begin{align}
v_\ell^{Z(i)} &  = \mathcal{O}_{1i}\left( -\frac{1}{2} + 2 s_W^2\right)
	- \frac{1}{\sqrt{g^2 + g^{\prime 2}}}\left(
		g_X \mathcal{O}_{3i} + \frac{1}{2} g_H \mathcal{O}_{2i} \right)\;,\\
v_u^{Z(i)} & = \mathcal{O}_{1i}\left(\frac{1}{2} - \frac{4}{3} s_W^2\right)
	+ \frac{1}{\sqrt{g^2 + g^{\prime 2}}}\left(
		g_X \mathcal{O}_{3i} + \frac{1}{2} g_H \mathcal{O}_{2i} \right)\;,\\
v_d^{Z(i)} & = \mathcal{O}_{1i}\left(-\frac{1}{2} + \frac{2}{3} s_W^2\right)
	- \frac{1}{\sqrt{g^2 + g^{\prime 2}}}\left(
		g_X \mathcal{O}_{3i} + \frac{1}{2} g_H \mathcal{O}_{2i} \right)\;,\\
a_\ell^{Z(i)} &  = - a_u^{Z(i)} = a_d^{Z(i)} = - \frac{\mathcal{O}_{1i}}{2}
	+ \frac{1}{\sqrt{g^2 + g^{\prime 2}}}\left(
		g_X \mathcal{O}_{3i} + \frac{1}{2} g_H \mathcal{O}_{2i} \right)\;,
\end{align}
where $i=1,2,3$ and $Z(1)\equiv Z$, $Z(2)\equiv Z'$,  $Z(3) \equiv A'$.
Using the relations between mixing matrix elements and mixing angles in
Eqs.~(2.6) to~(2.9) of Ref.~\cite{Huang:2019obt}, we can find the following relation
\begin{equation}
	\frac{2}{\sqrt{g^2 + g^{\prime 2}}}
	\frac{\left(g_X \mathcal{O}_{3i} + \frac{1}{2} g_H
	\mathcal{O}_{2i}\right)}{O_{1i}} = 1 - \frac{m_{Z(i)}^2}{m_{Z^{\rm SM}}^2} \; ,
\end{equation}
which can be used to simplify the above vectorial and axial couplings to obtain
\begin{align}
v_\ell^{Z(i)} & = -\mathcal{O}_{1i}\left( 1 - 2 s_W^2 - \frac{r_i^2}{2}\right)\;,\\
v_u^{Z(i)} & = \mathcal{O}_{1i}\left(1 - \frac{4}{3} s_W^2 - \frac{r_i^2}{2}\right)\;,\\
v_d^{Z(i)} & = -\mathcal{O}_{1i}\left(1 -\frac{2}{3} s_W^2 - \frac{r_i^2}{2}\right)\;,\\
a_\ell^{Z(i)} & = - a_u^{Z(i)} = a_d^{Z(i)} = -
\frac{\mathcal{O}_{1i}}{2}r_i^2\;,
\end{align}
where $r_i=m_{Z(i)}/m_{Z^{\rm SM}}$. In this form, it is obvious that the axial
couplings magnitude is expected to be smaller than the ratio of squared masses
$r_i^2$, which, {\it e.g.}, for $m_{Z(i)}= 1$~GeV is already close to 10$^{-4}$.
In contrast, the vectorial couplings have an $r_i$-independent part whose size
is controlled by the mixing matrix element $\mathcal{O}_{1i}$ that also
affects axial couplings.
Therefore, for sufficiently light $Z'$ and $A'$, the contributions from the axial couplings
are expected to be subleading.

The neutral current interactions induced by $W^{\prime(p,m)}$ are given by 
\be
{\cal L}\left(W'\right) = g_H \left(J^{p\, \mu} W_\mu^{\prime p} + {\rm H.c.} \right), 
\label{eq:wpinteraction}
\ee
where 
\be
J^{p\,\mu} = \frac{1}{\sqrt{2}} \left[ \overline{u_R} V_u^H  \gamma^\mu u_R^H 
                    +  \overline{d_R^H} \left(V_d^H\right)^{\dagger}  \gamma^\mu d_R
                    + \overline{\nu_R} V_\nu^H  \gamma^\mu \nu_R^H 
                    +  \overline{e_R^H} \left(V_e^H\right)^{\dagger}  \gamma^\mu e_R  \right] \; ,
\ee
with $V_u^H$, $V_d^H$, $V_\nu^H$ and $V_e^H$ being the new unitary mass rotation matrices for the fermions. 
The neutral current interactions induced by $Z$, $Z'$, $A'$ bosons can be found in Ref.~\cite{Huang:2019obt}.


\subsection{Free Parameters\label{sec:free_params}}

From the scalar potential we recognize 9 free parameters including
couplings and VEVs.
Of those 9 parameters, $\mu_H^2$ and $\mu_\Phi^2$ can be related to other
parameters using the minimization conditions according to
Eqs.~\eqref{eq:muh2min} and~\eqref{eq:muphi2min}, leaving only 7 free
parameters.
Furthermore, Eq.~\eqref{eq:mixH1} can be used to relate the parameters
$\lambda_H$, $\lambda_\Phi$ and $\lambda_{H\Phi}$ to the physical squared
masses $m_{h_1}^2$, $m_{h_2}^2$ and the mixing angle $\theta_1$ obtaining the
following relations
\begin{align}
\label{eq:lamHmassangle}
\lambda_H & = \frac{1}{2 v^{2}}\left(m^{2}_{h_1} \cos^{2}{\theta_{1} } + m^{2}_{h_2}
\sin^{2}{\theta_{1} }\right) \; , \\
\label{eq:lamPhimassangle}
\lambda_\Phi & = \frac{1}{2 v_\Phi^{2}}\left(m^{2}_{h_1} \sin^{2}{\theta_{1} } +
m^{2}_{h_2} \cos^{2}{\theta_{1} }\right) \; , \\
\label{eq:lamHPhimassangle}
\lambda_{H\Phi} & = \frac{1}{2 v v_\Phi}\left[\left( m^{2}_{h_2} - m^{2}_{h_1} \right) \sin{\left(2
\theta_{1} \right)}\right] \; .
\end{align}
Using the mass of $D$, Eq.~\eqref{eq:Dscalarmass}, we can express
$\lambda'_{H\Phi}$ in the form
\begin{equation}
\label{eq:lampHPhimassangle}
\lambda'_{H\Phi} = \frac{2 m^{2}_{D}}{v^{2} + v_{\Phi}^{2}} \; .
\end{equation}
Using this last expression and the mass of the charged Higgs of
Eq.~\eqref{eq:chargedHiggsmass}, we can write $\lambda'_H$ as
\begin{equation}
\label{eq:lampHmassangle}
\lambda'_H = \frac{2}{v^2} \left[
	\frac{m^{2}_{D} v_{\Phi}^{2}}{v^{2} + v_{\Phi}^{2}}
	- m^{2}_{H^\pm}
\right] \; .
\end{equation}
Finally, we can use Eq.~\eqref{eq:Wprimepmmass} to relate $v_\Phi$ to the $W'$ mass as
\begin{equation}
v_\Phi^2 = \frac{4 m_{W'}^2}{g_H^2} - v^2 \; .
\end{equation}
Using the expressions in this subsection allow us to trade six model
parameters with five physical squared masses and one mixing angle,
\begin{equation}
\lambda_H, \lambda_\Phi, \lambda_{H\Phi}, \lambda'_{H\Phi}, \lambda'_H, v_\Phi
\:\:\:\:\to\:\:\:\:
m^2_{h_1}, m^2_{h_2}, m^2_D, m^2_{H^\pm}, m^2_{W'}, \theta_1 \; .
\end{equation}
The remaining free parameters of the model are the heavy fermion masses
$m_{f^{\rm H}}$, the Stueckelberg mass $M_X$, and the gauge couplings $g_X$ and
$g_H$.
Considering that the mass of the Higgs, $m_{h_1}$, has already been
measured~\cite{Zyla:2020zbs}, we are left with a total of 8 free parameters
plus the masses for 12 heavy hidden fermions. The effects of heavy hidden fermions in complex scalar dark matter 
phenomenology in G2HDM was analyzed in~\cite{Dirgantara:2020lqy}.

\section{Constraints\label{sec:constraints}}

In this section, we examine the model 
using various constraints including the theoretical constraints of the scalar potential, 
electroweak precision data, dark photon physics and
Higgs measurements at LHC\@. 
DM constraints are presented separately in the next section.


\subsection{Theoretical Constraints\label{sec:theory_constraints}}

The theoretical constraints on the original G2HDM
were studied in Ref.~\cite{Arhrib:2018sbz}.
Here, we follow closely on the steps of that work
but remove the scalar triplet and related parameters from the original model.

{\it Vacuum Stability\/}:
To make sure the scalar potential is bounded from below,
the sum of all quartic terms in the scalar potential needs to be positive.
In the same way as Ref.~\cite{Arhrib:2018sbz} we use copositivity conditions
given by the following constraints
\be
\widetilde \lambda_H (\eta) \geq 0 \; , 
\;\;\; \lambda_\Phi \geq 0 \; , 
\;\;\; \widetilde \lambda_{H\Phi}(\xi) + 2 \sqrt{\widetilde \lambda_H (\eta) \lambda_\Phi}  \geq  0 \; ,
\label{eq:copos}
\ee
where $\widetilde \lambda_H (\eta) \equiv \lambda_H + \eta \lambda^\prime_H$ and
${\widetilde  \lambda_{H\Phi}}(\xi) \equiv \lambda_{H\Phi} + \xi \lphp $.
The conditions of Eq.~\eqref{eq:copos} have to be met for any value of $\xi$
and $\eta$ in the ranges $0 \leq \xi \leq 1$ and $-1 \leq \eta \leq 0$.

{\it Perturbative Unitarity\/}: It is required that the parameter space
remains within the perturbative limits.
To this end, here we compute the 2$\to$2 scalar scattering
amplitudes induced by the
quartic couplings.
The 2$\to$2 processes induced by vertices
from scalar cubic couplings and gauge interactions are suppressed
by large momentum exchange in their propagators~\cite{Arhrib:2018sbz}.

Perturbative unitarity requires
\begin{align}
&\vert \lambda_H \vert, \vert  \lambda_\Phi \vert  \leq  4 \pi \;,
 \vert \lambda_{H\Phi} \vert  \leq  8 \pi \;,
 \vert \lambda^\prime_{H\Phi} \vert \;, \vert \lambda_H^\prime \vert  \leq  8 \sqrt 2 \pi \; , \\
& \vert 2 \lambda_H \pm \lambda^\prime_H \vert \leq 8 \pi \;,
\vert \lambda_{H\Phi} + \lambda_{H\Phi}^\prime \vert \leq  8 \pi \; ,\\
& \left| ( \lambda_H + \lambda_H^\prime/2 +
\lambda_\Phi) \pm \sqrt{2\lambda^{\prime 2}_{H\Phi} +
( \lambda_H + \lambda_H^\prime/2 - \lambda_\Phi)^2} \right|  \leq  8 \pi \;,  \\
& \left| (5 \lambda_H - \lambda_H^\prime/2 + 3 \lambda_\Phi) 
\pm \sqrt{(5 \lambda_H - \lambda_H^\prime/2 - 3\lambda_\Phi)^2 
+ 2 (2 \lambda_{H\Phi} + {\lambda}'_{H\Phi} )^2} \right|  \leq  8 \pi \; .
\end{align}


\subsection{Electroweak Constraints\label{sec:EW_constraints}}

A comprehensive study on the electroweak precision constraints
in the original model has been preformed in Ref.~\cite{Huang:2019obt}.
We found that the $Z$ mass shift is the most stringent among all
the electroweak precision constraints.
In particular, for the parameter space of interest in our analysis, 
\emph{i.e.}, 
$g_H \sim g_X \ll 1$ and $(m_{W'}^2, M_{X}^2) \ll m_{Z^{\rm SM}}^2$, the $Z$ mass shift, 
$|\Delta m_Z| = |m_Z - m_{Z^{\rm SM}}|$,
can be estimated as follows
\be
\label{deltaMZ}
\left|\frac{\Delta m_Z}{m_{Z^{\rm SM}}} \right| \simeq \frac{5}{2} \frac{g_H^2}{g_Z^2} 
\left(1+\frac{7}{5} \frac{m_{W'}^2}{m_{Z^{\rm SM}}^2} - \frac{4}{5} \frac{M_{X}^2}{m_{Z^{\rm SM}}^2 } \right)^{-1} \; , 
\ee
where $g_Z = g/\cos \theta_W$ with $\theta_W$ being the Weinberg angle. 
Following the methodology of Ref.~\cite{Feldman:2006ce}, we can obtain the experimental uncertainty of the $Z$ mass as 
\be
\bigg[ \frac{\delta m_Z}{m_{Z^{\rm SM}}} \bigg]^{2}=
\bigg[ \frac{c_W^{-2}-2 t_W^2}{\delta m_{W}^{-1}  m_W}\bigg]^{2}
+ \; \frac{t_W^4(\delta \Delta r)^{2}}{4(1-\Delta r)^{2}} \; ,
\label{eq:uncerMZ}
\ee
where $t_W = \tan\theta_W$ and $\Delta r$ is the radiative correction.
Using the PDG values~\cite{Zyla:2020zbs} of $m_W \pm \delta m_W = 80.387 \pm 0.016$ GeV, 
$\Delta r \pm \delta \Delta r =  0.03652 \mp 0.00021 \pm 0.00007$ and
$\sin \theta_W = 0.22343$, and by requiring $|\Delta m_Z| < |\delta m_Z|$,
one obtains an upper bound on $g_H$ and $g_X$ 
\be
|g_X| \sim |g_H| \lesssim 0.006  \,\times \sqrt{1-\frac{7}{5} \frac{m_{W'}^2}{m_{Z^{\rm SM}}^2} + \frac{4}{5} \frac{M_{X}^2}{m_{Z^{\rm SM}}^2 } }\; .
\label{eq:eps2MZ}
\ee
%

\subsection{Dark Photon\label{sec:dark_photon}}

The light boson $A'$ can be treated as a dark photon and, therefore, dark
photon constraints have to be applied.
In particular, due to the vertex $A'\bar{\ell}\ell$,
where $\ell$ represents any charged lepton,
it is expected that for a sufficiently large coupling it should
be possible to observe the $A'$ resonance in the invariant mass distribution of
$e^+e^-$ and $\mu^+\mu^-$.
Dark photon experiments constrain the size of the coupling
via a parameter $\varepsilon_\ell$.
In the decay width $\Gamma\left(A' \to \ell\ell\right)$ the parameter
$\varepsilon_\ell$ appears as~\cite{Fabbrichesi:2020wbt}
\begin{equation}
\label{eq:Apdecay}
\Gamma\left(A' \to \bar{\ell}\ell\right) = \frac{\alpha}{3}\varepsilon_\ell^2 m_{A'}
\sqrt{1 - \mu_\ell^2}\left(1 + \frac{\mu_\ell^2}{2}\right),
\end{equation}
where $\mu_\ell = 2m_\ell/m_{A'}<1$ since this decay channel only opens for
$m_{A'} > 2 m_\ell$.
In the G2HDM, the parameter $\varepsilon_\ell$ at tree level is given by
\begin{equation}
\label{eq:epsilon}
\varepsilon_\ell = \frac{1}{2 s_W c_W}\sqrt{\left(v^{A'}_\ell\right)^2
	+ \left(a^{A'}_\ell\right)^2\left(\frac{1 - \mu_\ell^2}{1 + \mu_\ell^2/2}\right)}\;,
\end{equation}
where $v^{A'}_\ell$ and $a^{A'}_\ell$ are given in Sec.~2 of
Ref.~\cite{Huang:2019obt}.
From Ref.~\cite{Huang:2019obt} we know that $v_\ell^{A'}$ and $a_\ell^{A'}$
are the same for all the charged leptons and, thus, the only distinction in
$\varepsilon_\ell$ between different flavors of leptons comes from $\mu_\ell$.
As mentioned in Sec.~\ref{sec:gauge_mass}, for light enough $A'$ the
axial coupling will be negligible and $\varepsilon_\ell$ is expected to be nearly
independent of $\mu_\ell$, as is usually the case in models with dark photon.
It is important to mention that, since $Z'$ is also expected to be light,
the dark photon experimental limits can also be applied as above
with $A'\to Z'$ in Eq.~\eqref{eq:Apdecay} and $\{v_\ell^{A'},a_\ell^{A'}\} \to
\{v^{Z'}_\ell,a^{Z'}_\ell\}$ in Eq.~\eqref{eq:epsilon}.
However, since $A'$ is lighter by definition it is expected to be more
strongly constrained.

There are several experiments with reported stringent limits for
$m_{A'} > 1$~MeV~\cite{Aaij:2019bvg,Lees:2014xha,Batley:2015lha,Banerjee:2018vgk,
Riordan:1987aw,Blumlein:2011mv,Blumlein:2013cua,darkphotonexp}. 
The existing limits on $\varepsilon_\ell$ for a dark photon mass $m_{A'} > 1$~MeV
are displayed on the top pane of Fig.~10 in Ref.~\cite{Fabbrichesi:2020wbt}.


\subsection{Higgs Collider Data\label{sec:Higgs_data}}


\subsubsection{Higgs Boson Mass}

As aforementioned, $h_1$ is identified as the observed Higgs boson at the LHC\@.
In this analysis, we take the mass of Higgs boson as
$m_{h_1} = 125.10\pm0.14$~GeV~\cite{Zyla:2020zbs}.


\subsubsection{Higgs Decays into Diphoton}

The decay rate for $h_1 \to \gamma  \gamma $ is given by
\be\label{Ratehigg}
\Gamma( h_1 \to \gamma \gamma ) = \frac{1}{64 \pi} m_{h_1}^3 
\biggl\vert F_{\gamma\gamma}( W^\pm ) 
+ F_{\gamma\gamma}( H^\pm ) 
+ \sum_{{\rm Charged} \, f^{\rm SM}} F_{\gamma\gamma}( f^{\rm SM} ) 
+ \sum_{{\rm Charged} \, f^{H}} F_{\gamma\gamma}( f^{\rm H} )  
 \biggr\vert^2 \; ,
\ee
where~\footnote{
Note that both the charged Higgs and heavy fermion contributions in $h_1 \to \gamma\gamma$ 
were not handled properly in~\cite{Huang:2015wts}.}
\begin{align}\label{FWloopgg}
F_{\gamma\gamma} ( W^\pm ) = {}& 
\frac{-1}{16 \pi^2} \cdot e^2 \cdot g \cdot \frac{1}{m_W} \cdot \cos \theta_1 \cdot
\left[ 2 + 3 \tau_{W} + 3 \tau_{W} \left( 2 -  \tau_{W} \right) f (  \tau_{W} ) \right]
\, ,  \\
\label{FCHloopgg}
F_{\gamma\gamma} ( H^\pm ) = {}& \frac{-1}{16 \pi^2} \cdot e^2 \cdot g_{h_1 H^+ H^-} \cdot \frac{1}{ m^2_{H^\pm} } \cdot 
\left\{ \tau_{H^\pm} \left[ 1 -  \tau_{H^\pm} f( \tau_{H^\pm}) \right] \right\}  \; , \\
\label{FfSMloopgg}
F_{\gamma\gamma} ( f^{\rm SM} ) = {}& 
 \frac{1}{16 \pi^2} \cdot N_c \cdot e^2 Q^2_{ f^{\rm SM}}  \cdot
 \frac{4} {v} \cdot \cos \theta_1 
\cdot \left\{ \tau_{f^{\rm SM}} \left[ 1 + \left( 1 - \tau_{f^{\rm SM}} \right) f ( \tau_{f^{\rm SM}} ) \right] \right\}
\; ,  \\
\label{FfHloopgg}
F_{\gamma\gamma} ( f^{\rm H} ) = {}& 
 \frac{-1}{16 \pi^2} \cdot N_c \cdot e^2 Q^2_{ f^{\rm H}}  \cdot
 \frac{4}{v_\Phi}\sin \theta_1
\cdot \left\{ \tau_{f^{\rm H}} \left[ 1 + \left( 1 - \tau_{f^{\rm H}} \right) f ( \tau_{f^{\rm H}} ) \right] \right\}
\; . 
\end{align}
Here, and in what follows, we define $\tau_i = 4 \,m_i^2/m_{h_1}^2$ where $i$ 
indicates which particle is running inside the loop. 
$N_c$ is the color factor, 1 for leptons and 3 for quarks.
The coupling $g_{h_1 H^+H^-}$ corresponds to the
$h_1 H^+H^-$ vertex and is given by 
\be\label{gh1CHCH}
g_{h_1 H^+H^-} =  \left( 2 \lambda_H - \lambda^\prime_H \right) v \cos \theta_1 
- \left( \lambda_{H \Phi} + \lambda^\prime_{H \Phi} \right) v_\Phi \sin \theta_1 \; . 
\ee
The well-known loop function $f(x)$ is
\begin{equation}\label{fx}
f( x ) = \left\{ 
\begin{array}{lr}
	\arcsin^2 \left(\frac{1}{\sqrt{x}}\right)   &   (x \geq 1) \, , \\
- \frac{1}{4} \left[
	\ln \left( \frac{1 + \sqrt{1 - x} } {1 - \sqrt{1 - x} } \right) - i \pi
	\right]^2\quad\quad   & (x < 1) \, .
\end{array} \right.
\end{equation}

The signal strength parameter for the Higgs boson produced from the gluon-gluon fusion (ggH) 
can be obtained as 
\be
\mu^{\gamma\gamma}_{\rm ggH} =  \frac{\Gamma_h^{\rm SM}}{\Gamma_{h_1}} 
\frac{\Gamma ({h_1\to gg}) \Gamma ({h_1\to \gamma\gamma}) } 
{\Gamma^{\rm SM} ({h \to gg}) \Gamma^{\rm SM} ( {h \to \gamma\gamma} )} \; ,
\ee
where the superscript SM refers to the SM Higgs boson $h$. 
The decay width of $h_1$ into two gluons is given by~\cite{Huang:2015wts}

\be
\Gamma (h_1\to gg) = \frac{\alpha_s^2}{48 \pi^3} m_{h_1}^3  
\left| \sum_{q^{\rm SM}} F_{gg}(q^{\rm SM}) 
+ \sum_{q^{\rm H}} F_{gg}(q^{\rm H}) 
\right|^2 \; ,
\ee
where 
\begin{align}
F_{gg}(q^{\rm SM}) = {}&   \frac{ \cos \theta_1}{v}  \tau_{q^{\rm SM}} \left[ 1
+ \left( 1 - \tau_{q^{\rm SM}} \right) f ( \tau_{q^{\rm SM}} ) \right], \\
F_{gg}(q^{\rm H}) = {}& - \frac{\sin \theta_1}{v_\Phi}  \tau_{q^{\rm H}} \left[
1 + \left( 1 - \tau_{q^{\rm H}} \right) f ( \tau_{q^{\rm H}} ) \right] \; , 
\end{align}
with $q^{\rm SM}$ and $q^{\rm H}$ refer to the SM quarks and the new colored fermions.
The latest measurement of this signal strength is given by ATLAS as
0.96$\pm$0.14~\cite{Aad:2019mbh}.


\subsubsection{$h_1 \to$ {\rm SM} Fermions}

The decay width of Higgs boson to SM fermions is given by 
\be
\Gamma({h_1 \to f^{\rm SM} \bar{f}^{\rm SM}} ) = \frac{N_c}{8 \pi} \frac{ m_{h_1}m_{f^{\rm SM}}^2}{v^2} 
 \left( 1- \tau_{f^{\rm SM}}\right)^{3/2} \cos^2 \theta_1 \; .
\label{eq:h1ff}
\ee
The signal strength for ggH production is then given by
\be
\mu^{ff}_{\rm ggH} = \cos^2 \theta_1 \frac{\Gamma (h^{\rm SM})}{\Gamma_{h_1}} 
\frac{\Gamma ({h_1\to gg})} {\Gamma^{\rm SM}({h \to gg})} \; .
\ee
Note that this expression is independent of the fermion flavor and thus we
compare against the best measured signal strength given by the decay into
a pair of $\tau^+\tau^-$:
$\mu^{\tau\tau}_{\rm ggH} = 1.05^{+0.53}_{-0.47}$~\cite{Sirunyan:2018koj}.


\subsubsection{Invisible Higgs Decay}

If $m_{h_1} > 2 \,m_{W'}$, the Higgs boson can decay invisibly into a pair of $W^{\prime (p,m)}$. 
The decay width of $h_1 \to W^{\prime p} W^{\prime m}$ is given by
\be
\Gamma ({h_1 \to W^{\prime p} W^{\prime m}}) = \frac{g_H^4 \left( v \cos \theta_1 - v_\Phi \sin \theta_1 \right)^2 }{256 \pi} 
\frac{m_{h_1}^3}{m_{W'}^4} \left( 1- \tau_{W'} +\frac{3}{4}  \tau_{W'}^2\right) 
\sqrt{1- \tau_{W'} } \; .
\label{eq:invdecaywidth}
\ee
In our parameter choice, we will assume $2 \, m_{\nu^H} > m_{h_1}$ so that $h_1$ does not decay into a pair of $\nu^H$.
The branching ratio of invisible Higgs decay is then given by 
\be
{\rm BR} ({h_1 \to {\rm inv}}) = \frac{\Gamma({h_1 \to W^{\prime p} W^{\prime m}})}{\Gamma_{h_1}} \; .
\ee
Recently, the ATLAS collaboration reported the most stringent constraint on
the invisible decays of the Higgs produced via vector boson fusion.
Assuming that the Higgs boson production cross section is comparable to the
SM, the ATLAS collaboration set the limit ${\rm BR}({h_1 \to{ \rm inv}}) <
0.13$ at $95 \%$ C.L.~\cite{ATLAS:2020cjb}. 


\section{Dark Matter Constraints\label{sec:DMPheno}}

\subsection{Relic Density\label{sec:relic}}

The DM scenario presented here works similarly to the very well known WIMP DM\@.
The DM candidate $W^{\prime (p,m)}$ begins in thermal
equilibrium with the particle species in the hot primordial soup
in the early universe before starting to
freeze-out due to the expansion of the universe.
The Boltzmann equation allows us to determine the evolution of the DM density
and precisely determine the amount of relics that remain after freeze-out.
This evolution is heavily influenced by the back and forth annihilation
(creation) of pair of DM particles into (from) pairs of SM states in the early
universe and their number densities.
An excessive annihilation of DM into SM particles would result in very low
relic density while not having enough annihilation would leave an overabundant
DM\@.
In our model, couplings between SM and BSM states have to be suppressed to
minimize the effects on the precisely measured properties of the $Z$ and the
Higgs, given that these measurements are in good agreement with the SM\@.

\begin{figure}[tb]
	\includegraphics{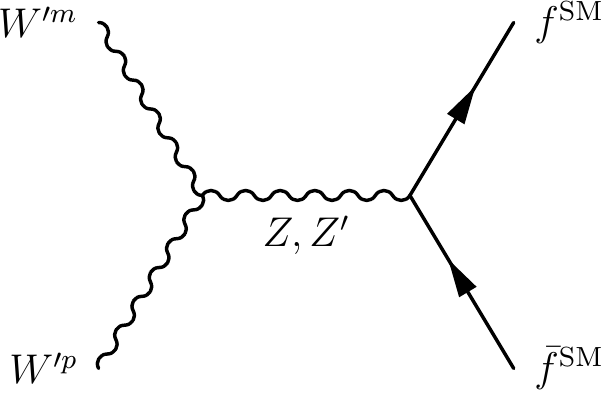}
	\caption{\label{fig:FeynAnnihilation} Dominant annihilation channels of
	the $W^{\prime \, (p,m)}$ DM candidate.}
\end{figure}

The main DM annihilation channels in our model are to pairs of SM fermions
mediated by $Z$ and $Z'$ as depicted in Fig.~\ref{fig:FeynAnnihilation}.
Other annihilation channels are also possible but are far more suppressed
compared to the channels just mentioned.
First, there is the $A'$ exchange diagram.
The $A'$ couplings to SM fermions are suppressed by combinations of
new gauge couplings in $v^{A'}_f$ and $a^{A'}_f$.
Similar to the case of $A'$, the $Z'$ couplings to the SM fermions are suppressed by its own 
$v^{Z'}_f$ and $a^{Z'}_f$. However, it is possible to have $m_{Z'}$ close to twice the mass of the
$W^{\prime (p,m)}$, resulting in an important contribution from resonant annihilations.
Secondly, we also have the $h_1$ and $h_2$ Higgs exchange diagrams. 
Their couplings to pairs of $W^{\prime(p,m)}$ and SM fermions are
suppressed by $g_H$ and light fermion masses $m_q/v$ respectively.
Finally, it is possible to have $t$-channel annihilation diagram via the exchange of a new heavy
fermion, $f^{\rm H}$, but this channel is suppressed by a factor of
$g_H$ on each of the two vertices of the diagram and by the mass of the heavy
fermion in the propagator.
The Feynman diagrams for the main annihilation processes as given in
Fig.~\ref{fig:FeynAnnihilation} can be computed straightforwardly and give rise 
to the following total cross section for  each final fermion pair
\begin{align}
\label{eq:totxsec}
\sigma(W^{\prime p}W^{\prime m}\to \bar{f}f) = {}&
	\frac{N_c g_M^2 g_H^2 m_{W'}^{2}}{72\pi s^2}
	\left(1 - \frac{4\, m_{W'}^2}{s}\right)^{1/2}
	\left(1 - \frac{4 \, m_f^2}{s}\right)^{1/2} 
	\left(\frac{s^{2}}{m_{W'}^{4}} + 20 \frac{s}{m_{W'}^{2}} + 12\right) \nonumber \\
& \times
\left(  \frac{\left(s - m_f^{2}\right)}{ 6 m_{W'}^{2}} \mathcal{V}_{+}  +  \frac{m_f^{2}}{2 m_{W'}^{2}} 
\mathcal{V}_{-}  \right) \, ,
\end{align}
where
\begin{align}
& g_M = \frac{\sqrt{g^2 + g^{\prime 2}}}{2} \, ,\qquad
\mathcal{\hat D}_k = 1 - \frac{m_k^2}{s} + i \frac{\Gamma_k m_k}{s} \, ,\\
& \mathcal{V}_{\pm} =
	\mathcal{O}_{21}^2\frac{v_f^2 \pm a_f^2}{|\hat{\mathcal{D}}_Z|^2}
	+ \mathcal{O}_{22}^2\frac{(v_f^{Z'})^2 \pm (a_f^{Z'})^2}{|\hat{\mathcal{D}}_{Z'}|^2}
	+ 2\mathcal{O}_{22}\mathcal{O}_{21}\left(v_f v^{Z'}_f \pm a_f a^{Z'}_f\right){\rm
	Re}\!\left(\frac{1}{\hat{\mathcal{D}}_Z \hat{\mathcal{D}}^*_{Z'}}\right)\; .
\end{align}

The annihilation mediated by the $Z$ is suppressed by a factor of ${\mathcal O}_{21}^2$
required to be small mostly by measurements on the decay width of the $Z$ and
the decay branching fractions that limit the $Z\to W^{\prime p} W^{\prime m}$
process.
As mentioned above, the channel mediated by the $Z'$ is also suppressed by
combinations of gauge couplings that make most of the size of $v^{Z'}_f$ and
$a^{Z'}_f$.
However, these suppressions are not as strong as the suppression in other
channels and when we include the effects from $Z'$ resonance it is possible to
bring the relic density to its expected value of $\Omega h^2 \sim 0.1$.

In our study, we will consider the measured value of $\Omega h^2 =
0.120\pm0.001$ as given by the Planck collaboration~\cite{Aghanim:2018eyx}.

\subsection{Direct Detection\label{sec:DD}}

Due to the small coupling between the DM candidate, $W^{\prime (p,m)}$, with
the SM-like states $h_1$ and $Z$ and the BSM states $h_2$, $Z'$ and $A'$ which
couple to the visible sector,
it is possible to have effects from DM scattering against nucleons in
detectors used in direct detection experiments.
In this case, we have to consider the elastic scattering between a DM
particle and the partons (both quarks and gluons) present in the nucleon.
The suppression of vertices works in the same way as in the annihilation
processes described in Sec.~\ref{sec:relic}, where $h_j$ mediated processes
are suppressed by a factor of $g_H^2 m_q^2/v^2$ in the cross section
with an additional $m_{h_j}^{-4}$ suppression from the propagator
since these interactions happen via $t$-channel.
Therefore, we are only left with the processes mediated by $Z$, $Z'$ and $A'$
in the $t$-channel.
Usually, for direct detection processes the momentum exchange is considered to be
very small and therefore $t$ is expected to be small as well.
This will result in amplitudes suppressed by the inverse squared of the mass
of the mediator meaning that the light states, $Z'$ and $A'$, will be less
suppressed.
In the approximation where the momentum exchange is smaller than the mass of
the mediator, we can write the interaction between DM and light quark $q$ as a contact
interaction given by
\begin{equation}
\label{eq:DDLeff}
	\mathcal{L}_{\rm CI-DD} = \sum_q  \sum_{i=2}^3 \frac{
	g_M g_H \mathcal{O}_{2i}v^{Z(i)}_q
		}{2 m_{Z(i)}^2}
		\left(W^{\prime p\, \mu} \partial_\nu W^{\prime m}_\mu - W^{\prime m\,
		\mu} \partial_\nu W^{\prime p}_\mu\right)\bar{q}\gamma^\nu q \; ,
\end{equation}
where $i=\{2,3\}$ corresponds to $Z(2) \equiv Z'$ and $Z(3) \equiv A'$ respectively.
It is worth noting that, as light as the mediators $Z'$ and $A'$ are, we can
still integrate them out thanks to the comparably small maximum momentum
transfer, $q_{\max}$.
The smallness of $q_{\max}$ is mostly due to $W'$ being small as well.
Consider $q_{\max} \sim 2 \, v_{\rm DM}\, m_{W'} m_A/(m_{W'} + m_A)$ with
$m_{W'} = 0.5$~GeV and $v_{\rm DM} = 10^{-3}$ c, and the target mass $m_A = 131$~GeV or $40$~GeV for 
xenon or argon target respectively.
In both cases $q_{\max} \sim O(1\ \text{MeV})$ while we expect $m_{A'} \gtrsim
O(10\ \text{MeV})$ due to constraints on dark photons.  Additionally, smaller
$m_{W'}$ results in even smaller $q_{\max}$.
Furthermore, for the axial part of the interaction with the quark, in the
small momentum exchange limit, only the space components of $\gamma^\nu$
remain but these components are suppressed by the $W^{\prime (p,m)}$ momentum
due to the derivatives $\partial_\nu W^{\prime (p,m)}$ in Eq.~\eqref{eq:DDLeff}~\cite{Arcadi:2017kky,Escudero:2016gzx}.
This, together with axial couplings that are comparably much smaller than the
vectorial ones results in an spin dependent cross section that is expected to
be several orders of magnitude smaller than the spin independent one.

From Eq.~\eqref{eq:DDLeff}, it is clear that the $A'$ mediated process is
expected to dominate the cross section unless
$|\mathcal{O}_{23}/\mathcal{O}_{22}| < |m_{A'}/m_{Z'}|^2$.
The case where both mediators participate equally is expected to happen only
through fine tuning of masses and mixings.
Therefore, we expect the cross section with the nucleons to be mostly mediated
by either $A'$ or $Z'$.
The elastic cross section between $W^{\prime (p,m)}$ and a
nucleon, $N$, is given by
\begin{align}
\label{eq:DDxsec}
\sigma^{\rm SI}_{W'N} & = 
	\sigma^{\rm SI}_{W'p}
	\frac{\sum_k \eta_k \mu_{A_k}^2 \left[Z_\text{atom} + (A_k - Z_\text{atom}) f_n/f_p\right]^2}
		{\sum_k \eta_k \mu_{A_k}^2 A_k^2} \; , \\
\label{eq:DDxsecproton}
\sigma^{\rm SI}_{W'p} & = \frac{\mu_{p}^2 g_M^2 g_H^2 \mathcal{O}_{2i}^2}{4
\pi m_{Z(i)}^4}f_p^2 \; ,
\end{align}
where $i = 2$ or $3$ depending on the dominant mediator according
to the discussion above,
$\mu_{p} = m_{W'}m_p/(m_{W'} + m_p)$ is the reduced DM-proton mass,
$\mu_{A_k} = m_{W'}m_{A_k}/(m_{W'} + m_{A_k})$ is the reduced DM-isotope
nucleus mass and $f_p$ and $f_n$ are effective couplings of the DM with
protons and neutrons, respectively.
The atomic number is $Z_\text{atom}$ and the isotope dependent variables $\eta_k$ and
$A_k$ are the abundance and mass number of the $k^\text{th}$ target isotope,
respectively.
Direct detection experiments usually report the number in
Eq.~\eqref{eq:DDxsec} assuming isospin conservation, \emph{i.e.}, $f_p = f_n$.
In that case, it is straightforward to see that the ratio of the sums over
isotopes reduces to 1 and $\sigma^{\rm SI}_{W'N}  = \sigma^{\rm SI}_{W'p}$.
However, in our case the couplings between quarks, $u$ and $d$, and the gauge
bosons, $Z'$ and $A'$, are all different due to their distinct SM charges leading to isospin
violation (ISV), \emph{i.e.}, $f_p \neq f_n$.
Following Refs.~\cite{Feng:2011vu,Yaguna:2016bga}, we can rescale the reported experimental
limit, $\sigma_\text{limit} \to \sigma_\text{limit}\times\sigma^{\rm
SI}_{W'p}/\sigma^{\rm SI}_{W'N}$ to account for ISV effects and use it to
limit $\sigma^{\rm SI}_{W'p}$ as given by Eq.~\eqref{eq:DDxsecproton}.
This rescaling depends on the mass of DM, the atomic numbers and the
ratio $f_n/f_p$, and, therefore, will be different for different points in the
parameter space.

To constraint the $W^{\prime (p,m)}$-proton cross section we will use the most
recent upper limits set by the experiments CRESST III~\cite{Angloher:2017sxg},
DarkSide-50~\cite{Agnes:2018ves} and XENON1T~\cite{Aprile:2019xxb}.

\subsection{Indirect Detection\label{sec:ID}}

Due to DM annihilation before freeze out happening through the resonance of an
otherwise suppressed channel,
the annihilation of DM in the present---after the shift in energy from the
early to the current Universe---loses the resonance resulting in a very low
annihilation cross section.
We have checked that the value of the total annihilation cross section in G2HDM
at the present time is of order $10^{-32}$ cm$^3 \cdot$s$^{-1}$ or below, much lower than
the canonical limits set for various channels by Fermi-LAT data~\cite{Ackermann:2015zua, Fermi-LAT:2016uux}.

\subsection{Mono-Jet\label{sec:monojet}}

\begin{table}[htbp]
\begin{tabular}{|c| c c c c c c c c |c|} 
\hline
 BP &  $m_{W'}$ & $M_X $  &  
 $m_{h_2}$ &  $m_{D}$& $m_{H^{\pm}}$ & $g_H$ & $g_X$ & $\theta_1$  & $\sigma_{p p \to W'^p W'^m j}^{\rm precut}$ \\ 
 & (GeV) & (GeV) & (TeV) & (TeV) & (TeV)& ($10^{-4}$) & ($10^{-4}$) & (rad)  & (fb)\\
\hline
\hline
   1
 & 1.0
 & 1.67
 & 0.79
 & 2.0
 & 2.12
 & 5.4
 & 3.2
 & 0.17
 & 3.0  \\
   2 
 & 0.17
 & 0.33
 & 2.9
 & 1.54
 & 1.64
 & 1.0
 & 0.4
 & 0.11
 & 3.8  \\ 
\hline 
\end{tabular}
\caption{\label{tab:BP}Parameters for the two benchmark points where the production cross
	section of the mono-jet signals with precuts are computed, as shown in the last column.}
\end{table}

The occurrence of energetic jets with large missing transverse momentum
has been searched by ATLAS~\cite{Aaboud:2017phn, ATLAS:2020wzf}
and CMS~\cite{Sirunyan:2017hci} collaborations.
However, the observed results are overall in agreement with the SM predictions
and only exclusion limits have been reported.
In our model, the process $ p p \to W'^p W'^m j$
can give rise to mono-jet events at the
LHC\@. 

To analyze the mono-jet signal at the LHC, we choose two benchmark points (BPs) 
which are shown in Table~\ref{tab:BP}\@.
These two BPs satisfy all theoretical, Higgs data and DM constraints.
We evaluate the signal process cross section using {\tt MadGraph}~5~\cite{Alwall:2011uj}
with precuts for jets $p_T^j > 30$~GeV and $|\eta_j| < 2.8$, and for the
missing transverse momentum $p_T^{\rm miss} > 100$~GeV.
It turns out the production cross sections with the precuts are about $3.0$~fb and $3.8$~fb for
BP 1 and 2 respectively, and dominated by $Z$ and $Z'$ mediated diagrams.
We generate $10^4$~events for the $ p p \to W'^p W'^m j$ process and
recast ATLAS mono-jet search~\cite{ATLAS:2020wzf} using {\tt MadAnalysis}~5~\cite{Dumont:2014tja}.
The most sensitive signal region is found to be in the window $E_T^{\rm miss} \in (700, 800)$~GeV
(the signal region EM7 in Ref.~\cite{ATLAS:2020wzf}).
The 95\% C.L. exclusion limits on the production cross section
are 400~fb and 680~fb for BP 1 and 2 respectively,
which are much larger than the signal expected from the model.
Therefore, the LHC with luminosity of 139~fb$^{-1}$ is not sensitive enough to
search for mono-jet events from this model.
However, the model can be probed by mono-jet searches at future hadron colliders
such as the High-Luminosity Large Hadron Collider (HL-LHC)~\cite{Apollinari:2017cqg},
the High-Energy Large Hadron Collider
(HE-LHC)~\cite{Benedikt:2018ofy} and the
Future Circular hadron-hadron Collider (FCC-hh)~\cite{Arkani-Hamed:2015vfh}.


\section{Results\label{sec:results}}


\subsection{\label{sec:methodology} Methodology}

The masses of the gauge bosons, $m_Z$, $m_{Z'}$ and $m_{A'}$, as well as the
constraints from Sec.~\ref{sec:constraints} are calculated through our own
fortran codes, except for the Higgs invisible decay which can be calculated
together with the DM constraints.
The DM constraints of Sec.~\ref{sec:DMPheno}, in particular relic density,
direct detection and indirect detection are calculated using
{\tt micrOMEGAs}~\cite{Belanger:2018ccd} and a set of model files generated by
{\tt FeynRules}~\cite{Alloul:2013bka}.
For the invisible decay branching ratio of the Higgs, we take advantage of
the use of {\tt CalcHEP}~\cite{Belyaev:2012qa} within {\tt micrOMEGAs} to calculate the
decay width along with the rest of the DM constraints just mentioned.

All the points outside the theoretical constraints of
Sec.~\ref{sec:theory_constraints} are simply rejected.
By the same token, the dark photon constraints are used to reject any
parameter combination of $\varepsilon_{e,\mu}$-$m_{A'}$ or
$\varepsilon_{e,\mu}$-$m_{Z'}$ located inside the currently excluded regions.
The rest of the constraints in Sec.~\ref{sec:constraints} are summed into a
total $\chi^2$ that also includes relic density and direct detection
cross section.
In the case of direct detection experiments, where a limit is reported at a
95\%~C.L.\ with null-signal assumption, we use a $\chi_\text{DD}^2$ of the form
\begin{equation}
\chi^2_\text{DD} = 4.61\times\left(
	\frac{\sigma_\text{theory}}{\sigma_\text{limit}}\right)^2,
\end{equation}
where the 4.61 factors allows $\chi_\text{DD}^2 = 4.61$ when we are exactly at
the 95\%~C.L.\ of this two-dimensional limit.\footnote{
For a one-tailed test, the 95\%~C.L.\ corresponds to $\Delta\chi^2 =
2.71$ and $\Delta\chi^2 = 4.61$ of a Gaussian distribution in one and two
dimensions, respectively.  For a two-tailed test, the same numbers correspond
to the 90\%~C.L.\ limit.}
In mass ranges where more than one limit exists we take the one with the
largest $\chi_\text{DD}^2$.
Note that, due to ISV, the largest $\chi^2$ for direct detection may not
correspond to the experiment with the smallest cross section.
Since direct detection limits are reported assuming $f_p = f_n$ in
Eq.~\eqref{eq:DDxsec}, it is possible for ISV ($f_p \neq f_n$) to produce some amount of
cancellation or enhancement of the limits depending on the atoms used in the
detector.
Calculating the cross section in the way described in Sec.~\ref{sec:DD} allows
us to account for ISV and the atoms used in different experiments.

In the case of Higgs invisible decay branching fraction, where a limit is reported
with a 95\%~C.L., the appropriate $\chi_\text{inv}^2$ is given by
\begin{equation}
\chi^2_\text{inv} = 2.71\times\left(
\frac{{\rm BR} ({h_1 \to {\rm inv}})}{0.13}\right)^2,
\end{equation}
where, similarly to direct detection, the 2.71 factor allows for $\chi^2_\text{inv}
= 2.71$ when our result is exactly at the reported 95\%~C.L. in the one-dimensional case.

\begin{table}[htbp]
\begin{tabular}{|c|c|} 
\hline
Parameter [units] & Range \\
\hline
\hline
$m_{h_1}$ [GeV]     & [124.26\,,\,125.94]  \\
$m_{h_2}$ [TeV]     & [0.3\,,\,10] \\
$m_{D}$ [TeV]       & [0.3\,,\,10] \\
$m_{H^{\pm}}$ [TeV] & [0.3\,,\,10] \\
$\theta_1$ [rad]         & [$-\pi/2$\,,\,$\pi/2$] \\
$\log_{10}(m_{W'}/\text{GeV})$ & [$-3$\,,\,2] \\
$\log_{10}(M_X/\text{GeV})$    & [$-3$\,,\,2] \\
$\log_{10}(g_H)$               & [$-6$\,,\,0] \\
$\log_{10}(g_X)$               & [$-6$\,,\,0] \\
\hline
$m_{f^{\rm H}}$ [TeV]                & 3 (fixed) \\
\hline
\end{tabular}
\caption{\label{tab:prior}
Ranges and values for the prior of the parameters used in this analysis.
All priors are taken as uniform inside their ranges for the parameters listed in
the table.
While the $m_{h_1}$ range in this table corresponds to the measured $\pm 6\sigma$
interval, {\tt emcee} takes care of sampling it according to the result of
the total $\chi^2$.
}
\end{table}

To sample the parameter space we use the affine invariant Markov Chain Monte
Carlo ({\tt MCMC}) ensemble sampler {\tt emcee}~\cite{ForemanMackey:2012ig} which
presents advantages such as fast calculation of parameter distributions in
several dimensions.
The initial prior and ranges of each parameter are contained in
Table~\ref{tab:prior}\@.
In particular, the parameters $m_{W'}$, $M_X$, $g_H$ and $g_X$ are scanned
in base-10 logarithmic scale.
This is mostly because we expect these parameters to be small but different
from zero and that their effects depend heavily on their orders of magnitude.
For the masses of the heavy fermions, we expect their contributions to be
heavily suppressed by the requirement that $m_{f^{\rm H}} =
\mathcal{O}(\text{1~TeV})$.
Therefore we consider all of them degenerated with a mass of $m_{f^{\rm H}} = 3$~TeV
putting them safely above any current search for heavy fermionic states.
The rest of the parameters are scanned with a uniform prior in linear scale.
To guarantee that our final distributions are independent of the initial
points we perform several small runs collecting $\mathcal{O}(10^4)$ points for
each run using $\mathcal{O}(100)$ walkers.
The initial points for the walkers are always allowed by theoretical and dark
photon constraints but otherwise random inside the prior.
After checking that the final distributions are consistent between different
runs, we perform a large scan with 300 walkers collecting 160,000 points after
burn-in and thinning.


\subsection{\label{sec:numericalresults} Numerical Results}

\begin{figure}
	\includegraphics[scale=\figscale]{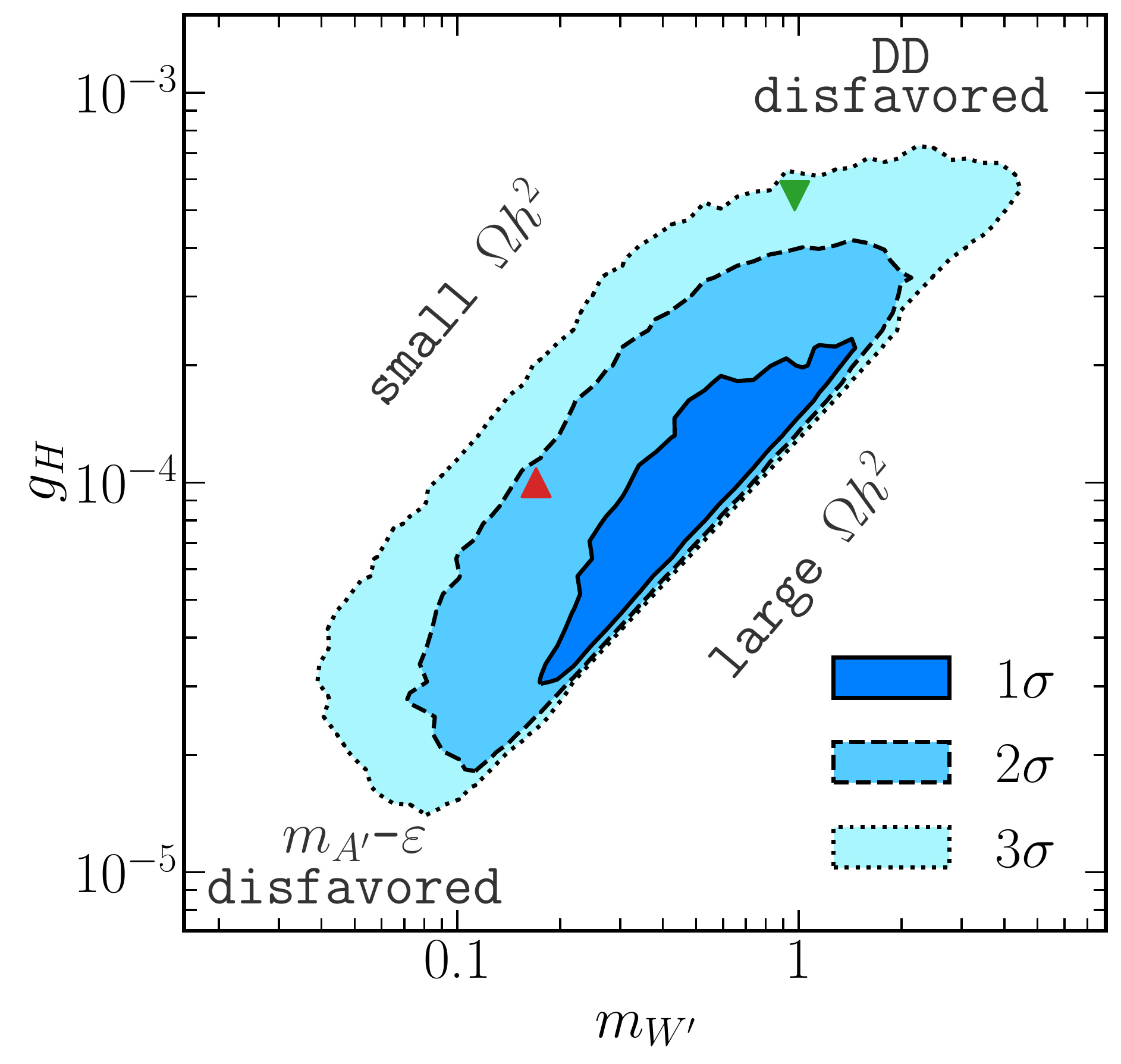}
	\includegraphics[scale=\figscale]{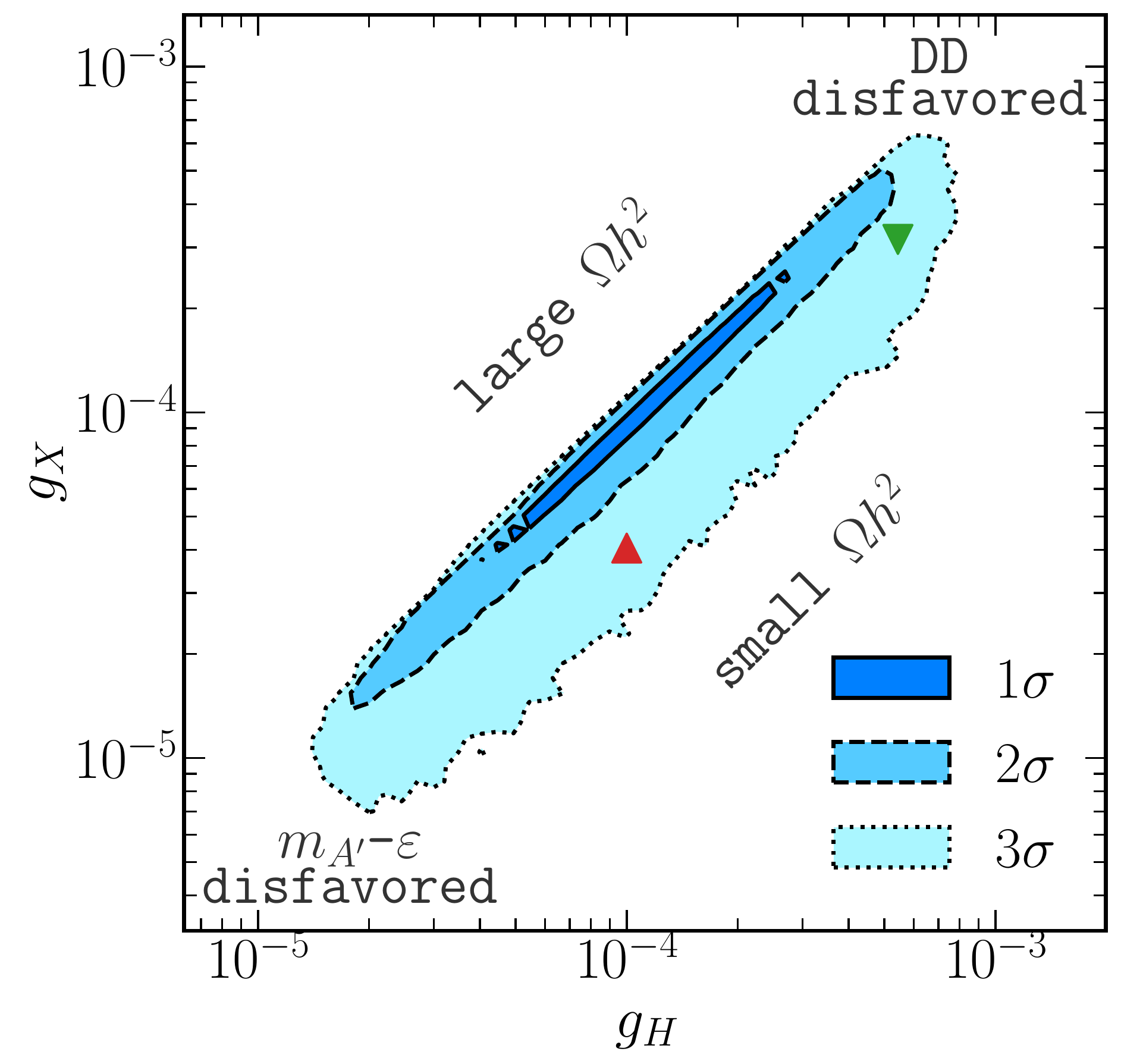}
	\caption{\label{fig:mwpgH_gHgX}
	The 1$\sigma$, 2$\sigma$ and 3$\sigma$ allowed regions (dark
	blue, medium blue and light blue, respectively) projected on the
	planes $(m_{W'}, g_H)$ (left) and $(g_H, g_X)$ (right).
	The solid, dashed and dotted black contours delimit the 1$\sigma$,
	2$\sigma$ and 3$\sigma$ regions respectively.
	The labels on empty zones represent the most relevant constraint in that
	region.
	The mono-jet benchmark points of Sec.~\ref{sec:monojet} are shown as a
	green down-triangle (BP 1) and a red up-triangle (BP 2).
	}
\end{figure}

\begin{figure}
	\includegraphics[scale=\figscale]{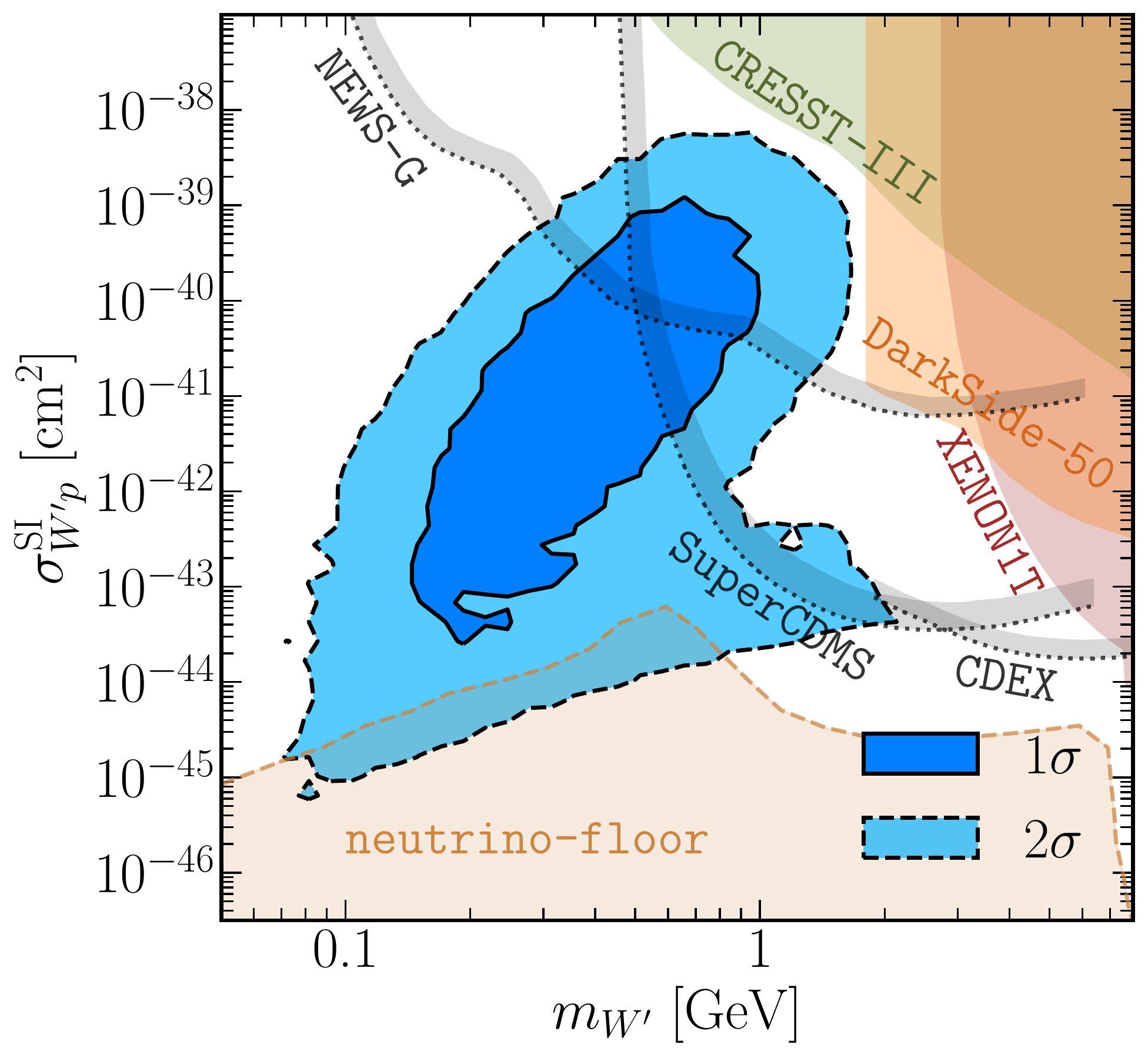}
	\includegraphics[scale=\figscale]{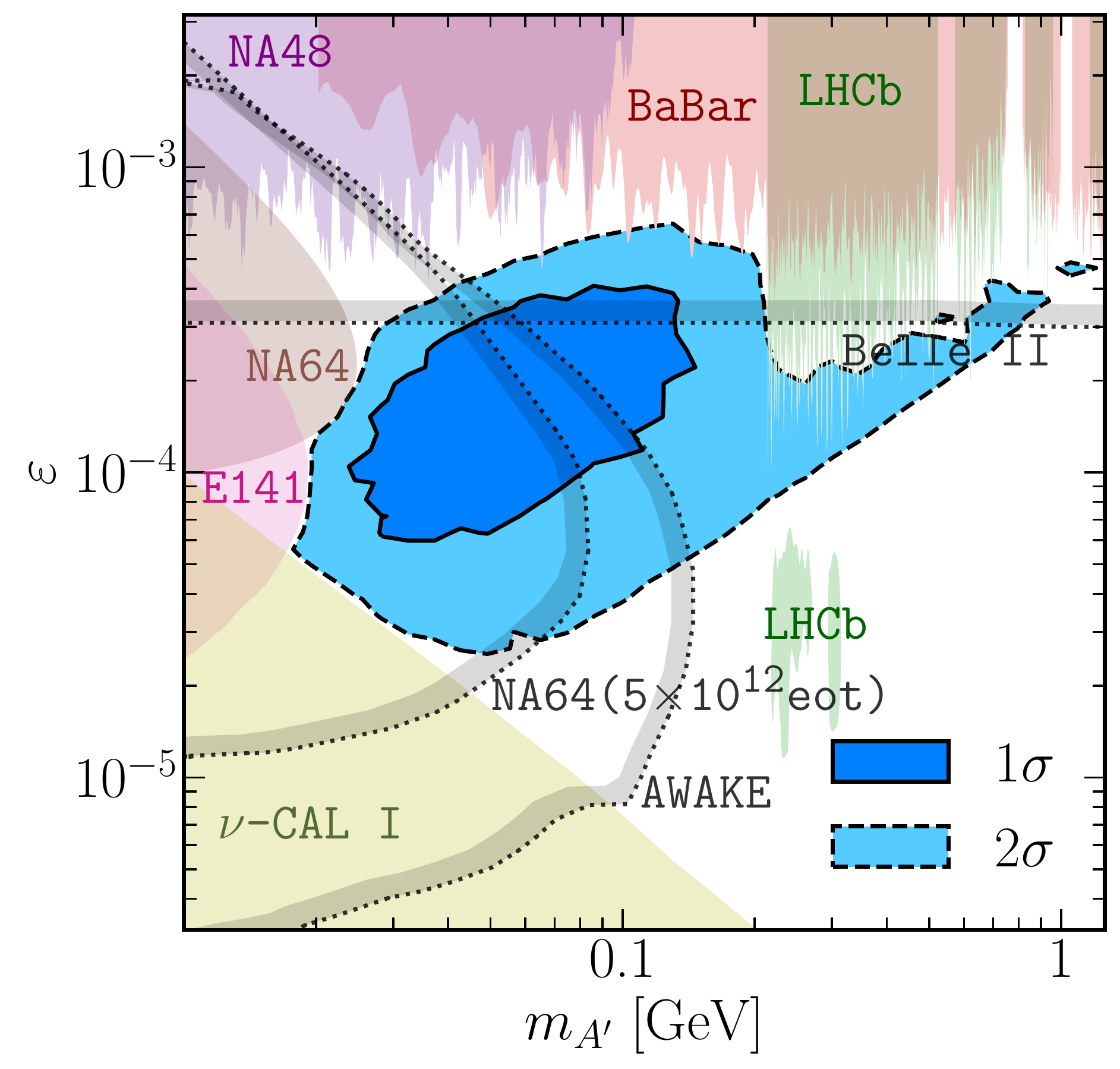}
	\caption{\label{fig:constraints}
	The 1$\sigma$ and 2$\sigma$ allowed contours
	projected on the DM mass, $m_{W'}$, vs.\ direct detection cross
	section (left) and dark photon mass, $m_{A'}$, vs.\ $\varepsilon$ coupling
	(right).
	The experimental excluded regions used in this study are shown as solid
	colored regions.
	Projected experimental limits are shown as dotted lines, with the
	direction of the exclusion marked in gray.
	In the case of direct detection (left), the limit set by the neutrino floor is
	shown as a dashed light orange line.
	}
\end{figure}

\begin{figure}
	\includegraphics[width=\textwidth]{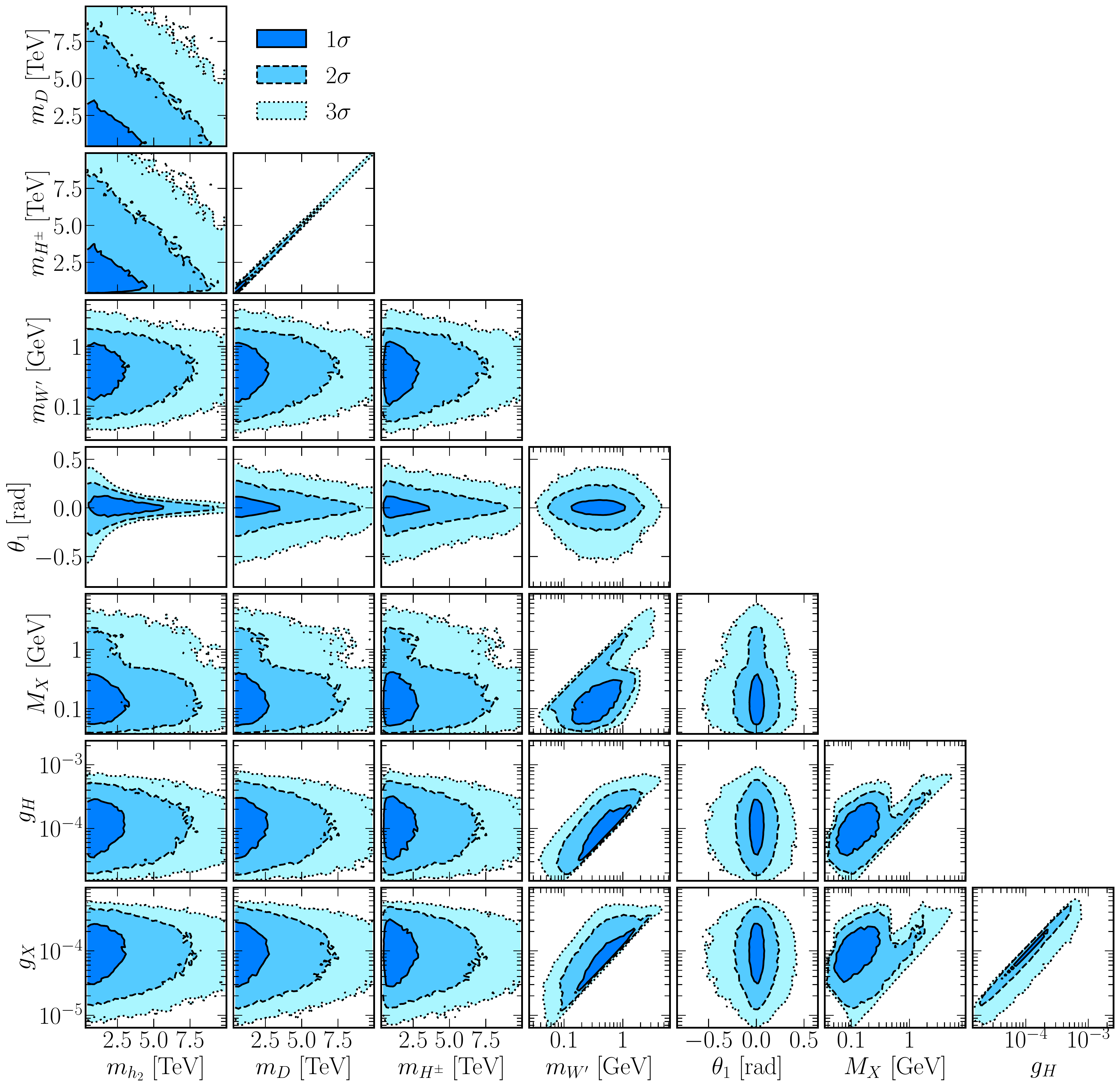}
	\caption{\label{fig:summary}
	Marginalized distributions in two dimensions for the parameters scanned in
	this study.
	The 1$\sigma$, 2$\sigma$ and 3$\sigma$ regions are marked in dark, light
	and lighter blue delimited by solid, dashed and dotted lines,
	respectively.
	}
\end{figure}

\begin{figure}
	\includegraphics[scale=\figscale]{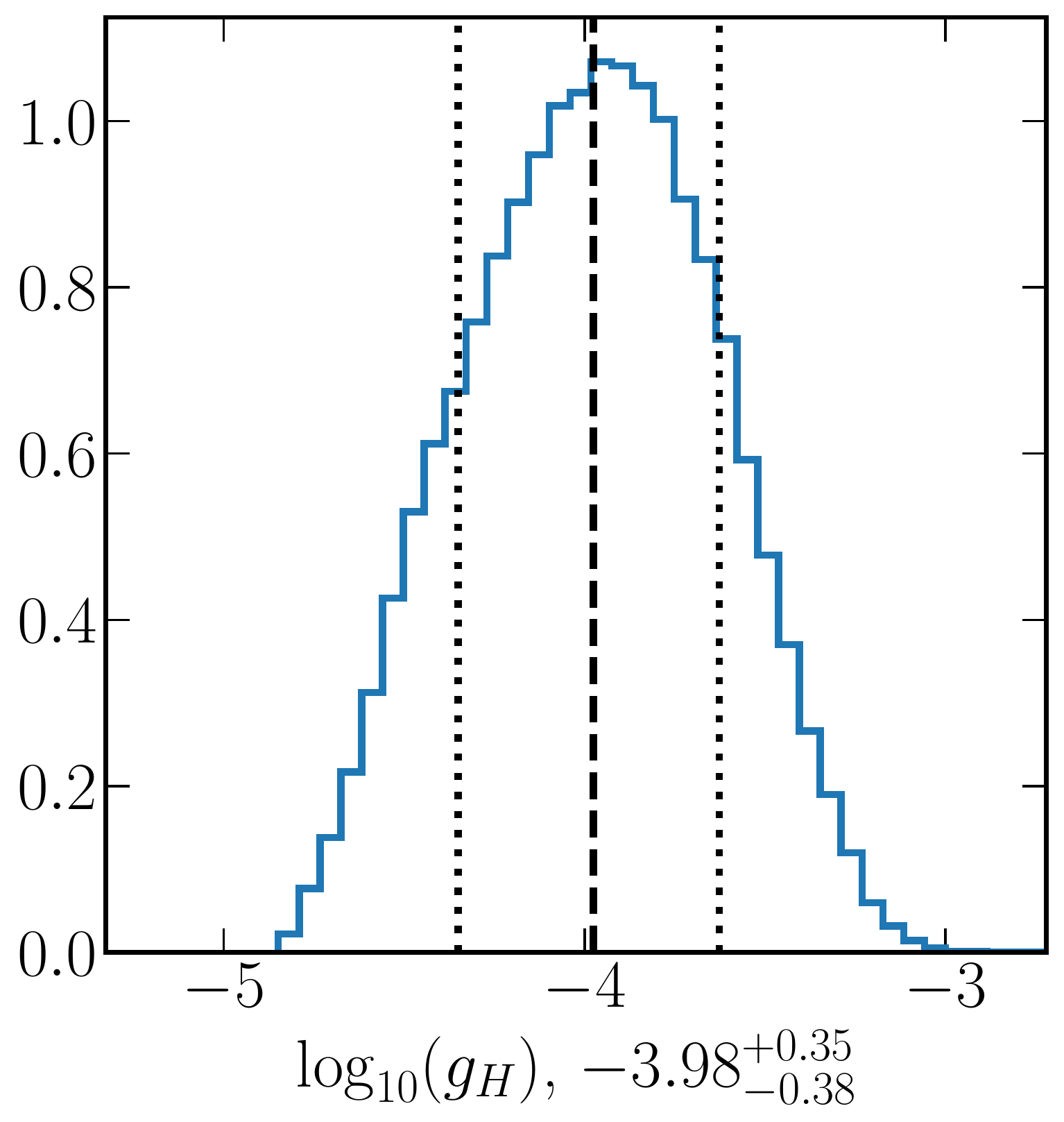}
	\includegraphics[scale=\figscale]{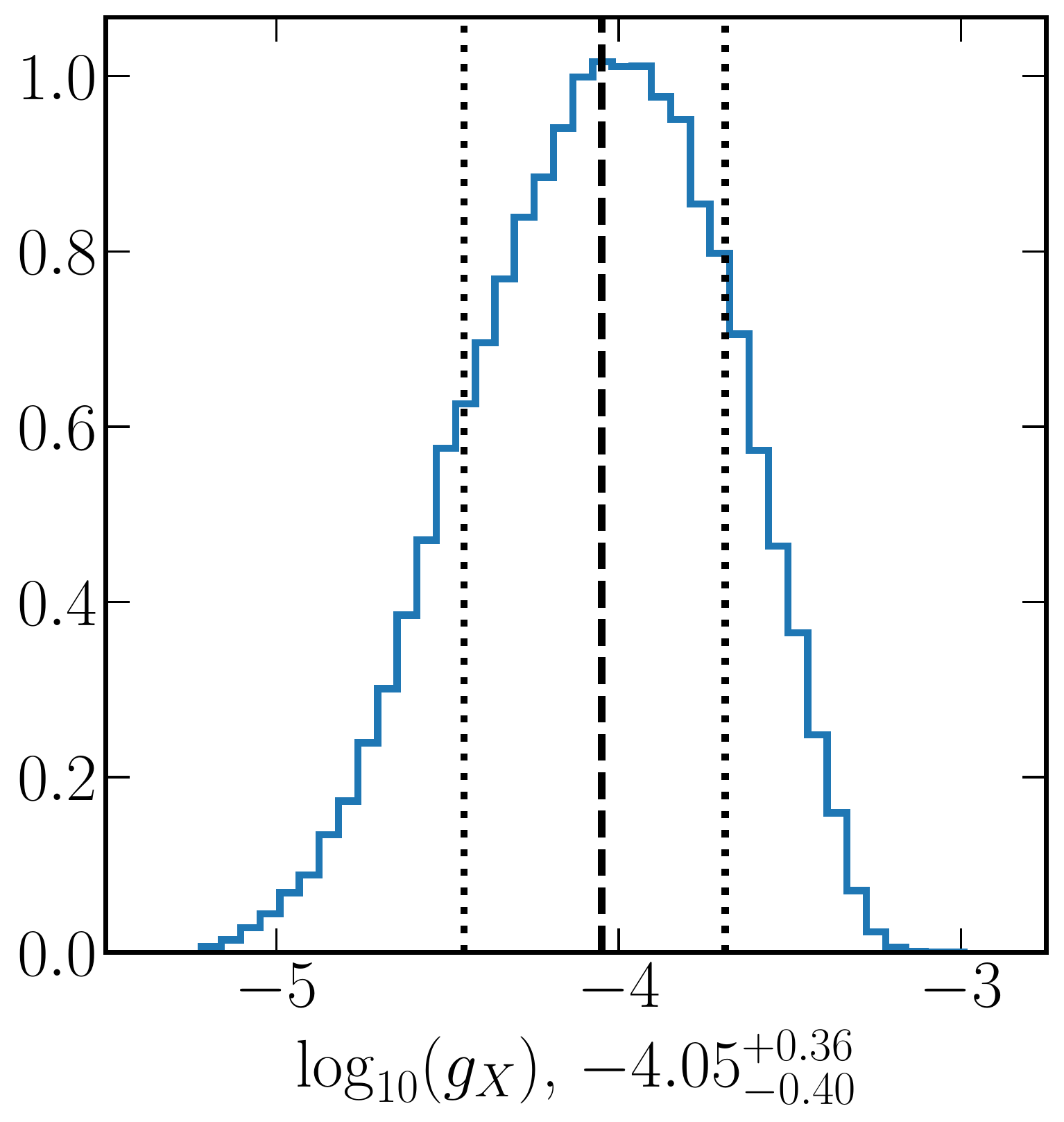}
	\includegraphics[scale=\figscale]{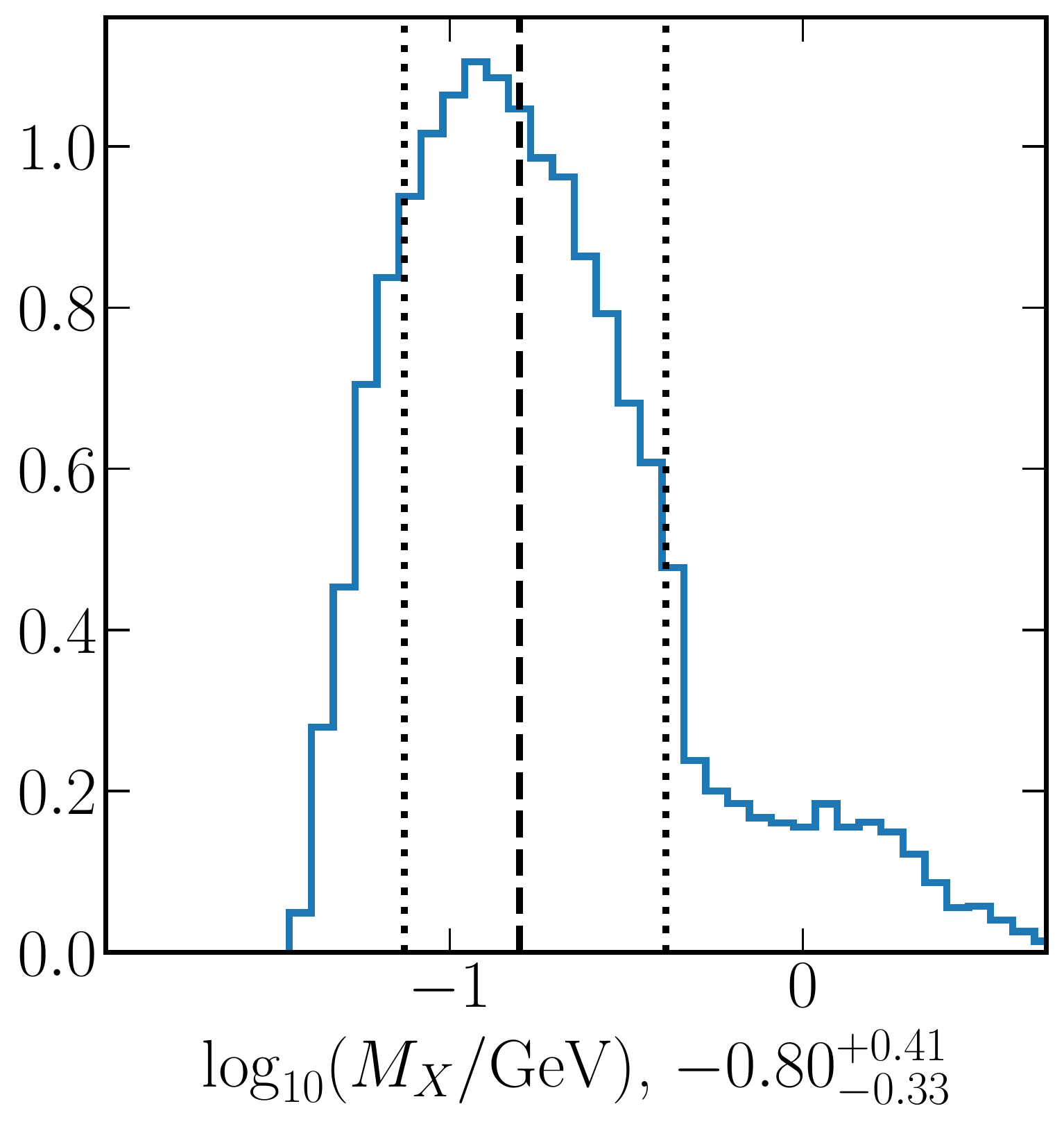}
	\includegraphics[scale=\figscale]{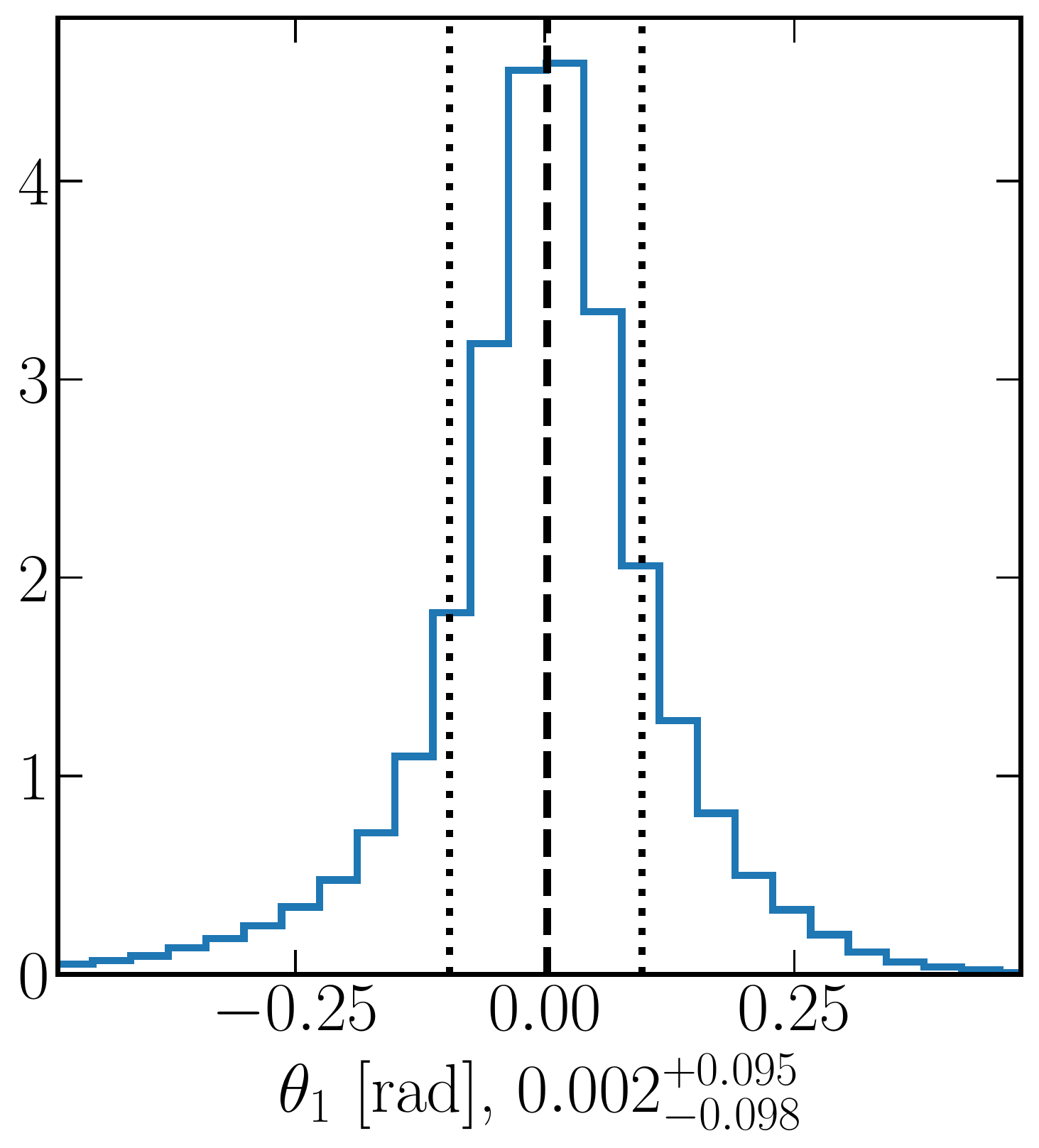}
	\includegraphics[scale=\figscale]{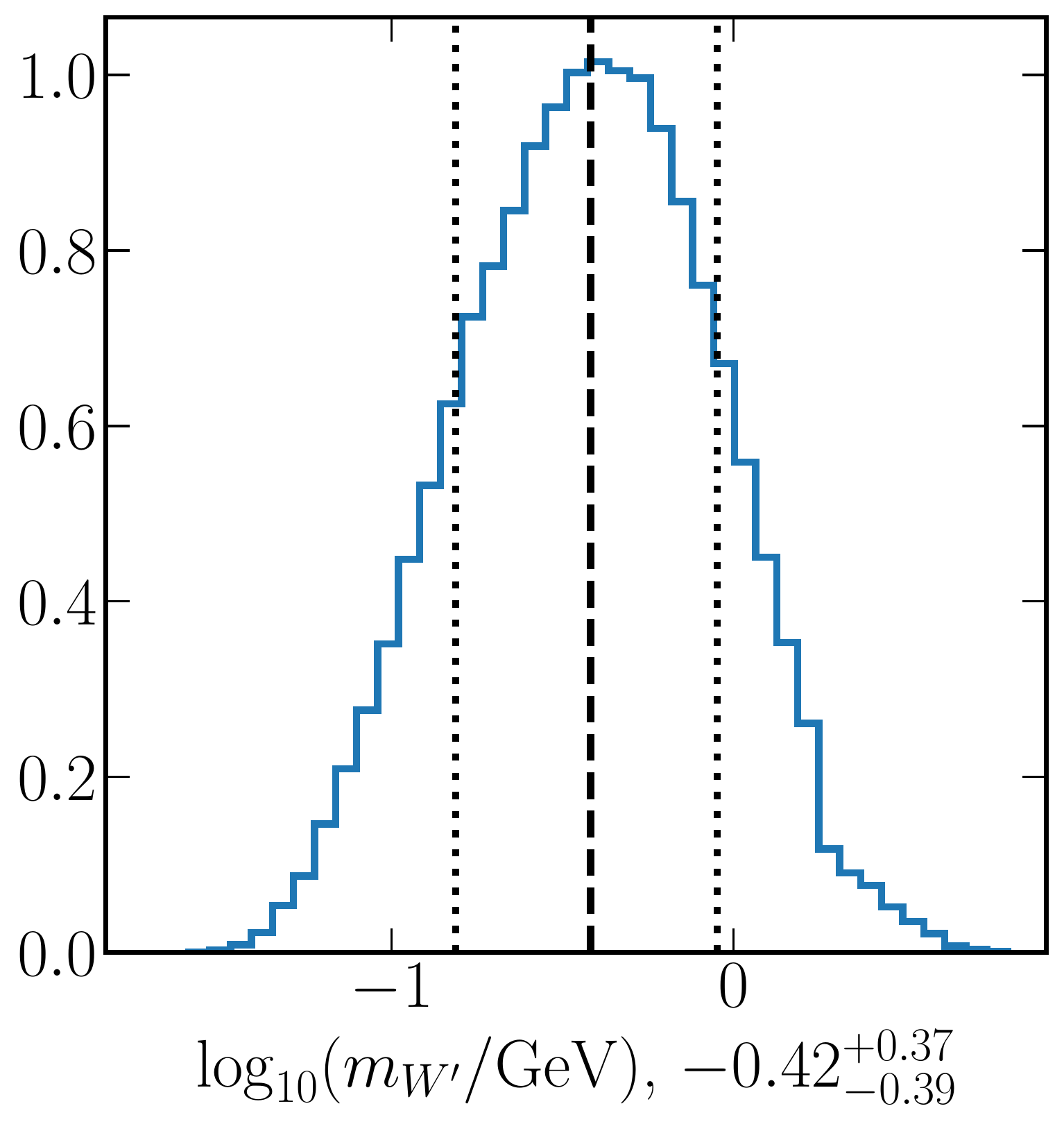}
	\caption{\label{fig:densities}
	Marginalized distributions of the most relevant parameters used in this
	study.
	Central values are indicated in the $x$-axis label.
	Since all the distributions are normalized, the $y$-axis values are only for
	reference.
	}
\end{figure}

We present the numerical results for visualization in
Figs.~\ref{fig:mwpgH_gHgX}--\ref{fig:densities}.
To follow the discussion below more smoothly, we suggest our readers to read the captions of these figures
first and then view and compare them in parallel.

The most notable feature of both panes of Fig.~\ref{fig:mwpgH_gHgX} is the 
band-shaped allowed region.
In the case of $(m_{W'}, g_H)$ plane shown in the left pane, the band is caused by the relation
between relic density and cross section, $\Omega h^2 \propto 1/\langle \sigma
v\rangle$.
Considering that we have $\sigma \propto g_H^2 m_{W'}^2/s^2$ from
Eq.~\eqref{eq:totxsec}, assuming $s\sim 4\, m_{W'}^2$ we have that $g_H^2
m_{W'}^2/s^2 \sim g_H^2/(16\,m_{W'}^2)$ resulting in $\Omega h^2 \propto
m_{W'}^2/g_H^2$.
This means that to keep a constant relic density, $m_{W'}$ and $g_H$ have to
keep a linear relationship as displayed in the left pane of
Fig.~\ref{fig:mwpgH_gHgX}.
Deviations from this band result in the relic density going either above or
below the value measured by the Planck satellite.

In the case of the right pane of Fig.~\ref{fig:mwpgH_gHgX}, the band can be
explained by the possibility of having a resonant annihilation of $W^{\prime
(p,m)}$ mediated by the $Z'$.
First note that Eq.~\eqref{eq:mapApprox} directly relates $m_{A'}$ and
$M_X$ and, as can be seen in the right pane of Fig.~\ref{fig:constraints}, $m_{A'}$ is required
to be mostly below 0.2~GeV due to the LHCb results, thus limiting also the
size of $M_X$.
Then, from Eq.~\eqref{eq:mzpApprox} we know that the term with $m_{W'}^2$
factor dominates over the term with $M_X^2$.
Finally, resonant annihilation is achieved for $m_{Z'} \approx 2\,m_{W'}$ meaning
$1 + 4g_X^2/g_H^2 \approx 4$ or $g_X^2/g_H^2 \approx 3/4$, resulting in the
band seen in the right pane of Fig.~\ref{fig:mwpgH_gHgX}.
Again, large deviations from this band result in too much or not enough
annihilation to achieve the correct relic density.
Due to this $g_X$-$g_H$ correlation, the
two-dimensional allowed regions in Fig.~\ref{fig:summary} for all the
distributions involving $g_X$ and $g_H$ have similar shapes.
Here it is important to mention that exact resonance, $m_{Z'} = 2\, m_{W'}$, would
result in too much annihilation and, therefore, in $\Omega h^2$ well below
Planck's measurement.

Besides their band-shaped tendency, both panes in Fig.~\ref{fig:mwpgH_gHgX}
are bounded in their top-right and bottom-left corners 
by the DM direct detection and
dark photon constraints, respectively.
We know that the direct detection cross section grows with $g_H^2$ as seen in
Eq.~\eqref{eq:DDxsecproton}, therefore, it is expected to see it setting an upper
bound on $g_H$.
Furthermore, as can be seen in the left pane of Fig.~\ref{fig:constraints},
direct detection experiments practically create a wall that limits the size of
$m_{W'}$ from above.
The effects of this limit are reflected in the upper bound of $m_{W'}$ in the
left pane of Fig.~\ref{fig:mwpgH_gHgX}.
In the case of the region disfavored by dark photon searches, this is mostly
due to the $\nu$-CAL I experiment limiting $\varepsilon$ from below as seen in the
right pane (olive green shaded zone) of
Fig.~\ref{fig:constraints}.
The $\varepsilon$ coupling limit is passed to $g_H$ through the vectorial and
axial couplings $v_f^{A'}$ and $a_f^{A'}$ that depend on it, resulting on the
lower limit on $g_H$ that can be seen in both panes of
Fig.~\ref{fig:mwpgH_gHgX}.

In the left pane of Fig.~\ref{fig:constraints} we show the allowed region projected on
the ($m_{W'}$, $\sigma_{W'p}^{\rm SI}$) plane,
where $\sigma_{W'p}^{\rm SI}$ is the cross section for spin-independent scattering on a proton.
The dark (light) blue shaded zone represents the $1\sigma$ ($2\sigma$)
allowed region.
The current DM direct detection measurements from
CRESST III (green)~\cite{Angloher:2017sxg},
DarkSide-50 (orange)~\cite{Agnes:2018ves} 
and XENON1T (brown)~\cite{Aprile:2019xxb}
constrain the DM mass to remain below $\sim 2$~GeV.
A small part of the $2 \sigma$ allowed region lies below the neutrino floor (light orange),
where the coherent neutrino-nucleus scattering would dominate over any DM signal.
Additionally, we show that experiments in the near future such as
NEWS-G~\cite{Battaglieri:2017aum}, SuperCDMS~\cite{Agnese:2016cpb} and CDEX~\cite{Ma:2017nhc}
can further probe our allowed parameter space, in particular for
$m_{W'} \gsim 0.3$~GeV with NEWS-G and down to $\sigma_{W'p}^{\rm SI} \sim
10^{-44}$~cm$^2$ with SuperCDMS and CDEX\@.

The right pane in Fig.~\ref{fig:constraints} shows the allowed region projected on
the ($m_{A'}$, $\varepsilon$) plane with the coupling $\varepsilon \equiv \varepsilon_\ell$.
Various experimental limits from dark photon searches are displayed in color shaded zones including
LHCb (green)~\cite{Aaij:2019bvg}, BaBar (pink)~\cite{Lees:2014xha},
NA48 (purple)~\cite{Batley:2015lha}, NA64 (light brown)~\cite{Banerjee:2018vgk},
E141 (magenta)~\cite{Riordan:1987aw} and $\nu$-CAL I (light green)~\cite{Blumlein:2011mv, Blumlein:2013cua}.
The dilepton searches at the LHCb, BaBar and NA48 put upper limits of
$\varepsilon \lesssim 10^{-3}$ for $m_{A'} \gsim 0.03$~GeV, especially LHCb
which sets a strong limit on $\varepsilon$ at $0.2$~GeV $< m_{A'} < 0.5$~GeV
causing a concave region
in the $2 \sigma$ allowed region at this mass range.
We note that this concave region due to LHCb corresponds to
the concave region at $(m_W', \sigma_{W'p}^{\rm SI}) \sim (1 \rm\,{GeV},
10^{-42}\, \rm{cm^2})$ in the left pane of the same figure. 
The LHCb long lived dark photon search constraints~\cite{Aaij:2019bvg} are also shown by the two isolated green shaded islands around
$\epsilon$ equals $2 \times 10^{-5}$.
On the other hand, the beam dump experiments NA64,
E141 and $\nu$-CAL I close the available space for smaller $\varepsilon$ and lighter
$m_{A'}$ setting lower bounds of $m_{A'} > 0.02$~GeV
and $\varepsilon \gsim 2\times 10^{-5}$.
The lower limit on $\varepsilon$ for $m_{A'} > 0.05$ GeV
is due to the DM relic density measured by the Planck experiment.
Interestingly, our final allowed region is located in
the gap between the beam-dump and the collider based experiments,
an area of special interest for future dark photon searches.
For example, Belle-II~\cite{Kou:2018nap} with a luminosity of
$50\ {\rm ab}^{-1}$ can probe $\varepsilon$ down to $2 \times 10^{-4}$, 
the next upgrade of NA64~\cite{NA64:2018} can cover
$10^{-5} \lsim \varepsilon \lsim 10^{-3}$ and $m_{A'} \lsim0.08$~GeV
by reaching $\sim 5 \times 10^{12}$ electrons-on-target (abbreviated by eot in the figure)
and Advanced WAKEfield Experiment (AWAKE) run 2~\cite{Caldwell:2018atq}
can reach $m_{A'}$ up to $0.15$~GeV with $10^{16}$
electrons-on-target with an energy of 50~GeV.
These limits are shown explicitly in the right pane of
Fig.~\ref{fig:constraints} as dotted lines with the side of the exclusion in
gray.
In the future, with access to high energy electron-proton colliders, AWAKE may
reach 1~TeV for the electrons, extending $m_{A'}$ up to
0.6~GeV~\cite{Caldwell:2018atq} and dark photon searches at LHeC and
FCC-he~\cite{DOnofrio:2019dcp} may even cover our entire allowed parameter
space.

We present two-dimensional projections of the allowed region
for our BSM parameters in Fig.~\ref{fig:summary}.
The dark, light and lighter blue zones indicate
the $1 \sigma$, $2 \sigma$ and $3 \sigma$ allowed regions respectively. 
One important thing to note is that the $m_{h_2}$, $m_{D}$ and
$m_{H^{\pm}}$ masses show an apparent upper bound at 1$\sigma$ and 2$\sigma$.
This upper bound actually depends on the maximum value chosen for the prior of
these three parameters and has no physical meaning.  In the case of $m_{h_2}$,
the apparent limit is due to the reduced $\theta_1$ for large $m_{h_2}$ seen
in the $(m_{h_2},\theta_1)$ subfigure while for $m_{D}$ and $m_{H^{\pm}}$ it is
due to their near degeneracy shown in the $(m_{D}, m_{H^{\pm}})$ subfigure.  We
have checked that changing the maximum scanned value for these three
parameters does not change the rest of the distributions, except for
$\theta_1$ where, understandably, larger $m_{h_2}$ sharpens the peak at
$\theta_1 = 0$ where $h_1 = h_{\rm SM}$ exactly.
The near degeneracy for $m_{D}$ and $m_{H^{\pm}}$ is due to their
comparably small mass squared difference
$m_D^2 - m_{H^\pm}^2 = v^2 (\lambda'_{H\Phi} + \lambda'_H)/2$.
Given that $(\lambda'_{H\Phi} + \lambda'_H)/2$ is bounded by unitarity
constraints, the mass squared difference is expected to remain
$\mathcal{O}(v^2)$ or less, meaning
that as $m_{D}$ and $m_{H^\pm}$ grow away from $v$ their proportional
difference rapidly grows smaller.
Their near degeneracy is also noticeable in all their two-dimensional
distributions since the distributions become nearly identical as $m_{D}$ and
$m_{H^\pm}$ grow larger.

Another interesting feature is that the charged Higgs mass
distributions reveal the presence of a lower limit around
$m_{H^\pm}\approx 400$~GeV where the contours show that the distribution falls rapidly. This is
due to the constraint from the Higgs decays into diphoton as shown in Eq.~\eqref{Ratehigg}.
Moreover,
due to the relation between $M_X$ and
$m_{A'}$ that can be inferred from Eq.~\eqref{eq:mapApprox},
the distributions in the $(M_X,g_X)$ and $(M_X,g_H)$ subfigures 
(second column from the right in (Fig.~\ref{fig:summary})) are close to the
distribution shown for the dark photon constraint in the right pane
of Fig.~\ref{fig:constraints}.

We present the marginalized one-dimensional distributions for the most
relevant parameters in Fig.~\ref{fig:densities} together with their central
values and their 1$\sigma$ intervals.
As mentioned before, there is a relation between $g_X$ and $g_H$ which is seen
again in their marginalized distributions in the top two panes since both peak
close to $10^{-4}$ and have very similar 1$\sigma$ intervals.
The distribution for $\log_{10}(M_X/\text{GeV})$ in the middle-left pane shows the effects of the LHCb
constraint around $\log_{10}(M_X/\text{GeV})\approx 0$ and peaks just above $-1$.
Because of the precise measurements on the Higgs properties at the LHC, 
the mixing angle between $h_1$ (identified to be the observed 125 GeV Higgs) 
and $h_2$, $\theta_1$ is found to have a 1$\sigma$ interval of $(-9.8,9.5)\times 10^{-2}$~rad
as shown in the middle-right pane of Fig.~\ref{fig:densities}.
For the distribution of the DM mass $m_{W'}$ in the bottom pane, we have the 1$\sigma$ interval for
$\log_{10}(m_{W'}/\text{GeV})$ between $-0.81$ and $-0.04$ which corresponds to
$m_{W'}\in [0.15, 0.91]$~GeV peaking at $m_{W'} \approx 0.39$~GeV.

After the above long discussions of our numerical analysis, perhaps a high level summary is useful.
\begin{itemize}

\item
We have focused our numerical analysis on the parameter space of the model that can lead to 
a sub-GeV $W^{\prime \, (p,m)}$ DM with a mass range of MeV$-$GeV.

\item
The viable domains of the parameter space are summarized in Fig.~\ref{fig:summary},
where the 1--3$\sigma$ contours allowed by all existing experimental constraints 
are shown for any combination of two of the 8 free parameters while the rest of the parameters are marginalized.  
All the masses of the new heavy fermions in the model, required by anomaly
cancellation, are set at 3 TeV\@.
For the two cases of $(m_{W^\prime}, g_H)$ and $(g_H, g_X)$, more detailed information on the boundaries of the contours 
due to the different constraints imposed are exhibited in Fig.~\ref{fig:mwpgH_gHgX}. 

\item
It is both clear and exciting to see from the allowed $1\sigma$ and $2\sigma$ contours on the plane 
of $m_{W^\prime}$ versus $\sigma^{\rm SI}_{W^\prime p}$ displayed in the left pane of Fig.~\ref{fig:constraints} 
that future experiments like NEWS-G and SuperCDMS can put more stringent constraints on the model. 
Indeed, about 1/3 of the current viable parameter space in the 
$(m_{W^\prime}, \sigma^{\rm SI}_{W^\prime p})$ plane would be facing challenge.
On the other hand, the region where the projected CDEX sensitivity can
reach is already disfavored in G2HDM\@.
Note that a small portion of the $2 \sigma$ allowed parameter space in the 
$(m_{W^\prime}, \sigma^{\rm SI}_{W^\prime p})$ plane is overlapping with the neutrino floor.

\item
For the allowed contours of dark photon coupling $\varepsilon$ and mass $m_{A^\prime}$ exhibited in the right pane of Fig.~\ref{fig:constraints},
future upgraded NA64 experiment with $5 \times 10^{12}$ electrons-on-target,
proof-of-principle experiment AWAKE run 2 with $10^{16}$ electrons-on-target, 
and next generation B-factory experiment Belle II would probe more than 1/2 of the current viable domain in the $(m_{A^\prime}, \varepsilon)$ plane.

\item
The correlation between the DM and dark photon physics and their constraints in G2HDM 
exhibited in the left and right panes of Fig.~\ref{fig:constraints} 
is rather novel and interesting. Indeed any non-abelian vector DM is likely accompanied by at least one extra neutral gauge boson 
which can play the role of $Z'$ or $A'$. 
Thus considerations of experimental constraints for both DM and dark photon physics must be taken into account mandatory.
Although our analysis is performed in the context of a specific model, we expect some of the features obtained in this work may be generic for 
any low mass non-abelian vector DM with a dominated vector portal of neutral gauge boson communicating to the SM sector. 
Perhaps we may be entertained by {\it nature} revealing to us not only a low mass dark matter $W^\prime$ 
but also a dark photon and a dark $Z^\prime$ with nearby masses. These are the three gauge bosons associated with the dark 
$SU(2)_H \times U(1)_X$ sector, in mirror with the visible SM gauge group $SU(2)_L \times U(1)_Y$.
Due to the minuscule couplings of $g_H$ and $g_X$, direct detections of the dark matter, dark photon and dark $Z^\prime$ 
(as well as other new particles introduced in G2HDM) at colliders
would belong to the lifetime frontier and high luminosity/energy frontier at the upgrade of LHC and future colliders.

\end{itemize}

\clearpage

\section{Conclusion\label{sec:conclusion}}

In this work, we studied a simplified version of the G2HDM~\cite{Huang:2015wts}.
In previous works, it was demonstrated that the original G2HDM successfully
explains dark matter while keeping a parameter space consistent with
theoretical and experimental expectations~\cite{Arhrib:2018sbz,Huang:2019obt,Chen:2019pnt}.
The simplifications considered in this study reduced the size of the scalar
potential and the parameter space by removing the scalar $SU(2)_H$ triplet.
Thankfully, the absence of this triplet does not affect the $h$-parity and
another dark sector particle, the $W^{\prime (p,m)}$, steps up as a DM candidate.
For the properties of $W^{\prime (p,m)}$ we settled on exploring the
sub-GeV mass range.
Interestingly, in this region the extra vector states $Z'$ and $A'$ happen to
have relevant roles as mediators for the annihilation and direct detection
processes of the DM\@.
In particular, reaching the appropriate amount of annihilation to have the
correct relic density would have been difficult if not impossible without the
channel mediated by the $Z'$.

We started with the usual theoretical checks on the scalar potential ensuring
that the minimum is stable and that the couplings remain unitary at tree level.
Given that the LHC has been closing in on the detailed properties of the 125 GeV Higgs, we
check that our scalar sector provides a particle, $h_1$, that matches the
mass and decays that have been measured.
The same can be said in the case of the gauge sector and the $Z$ with its
properties already very well measured at LEP\@.
In the case of the light $Z'$ and $A'$ we constrained their interactions with
SM leptons by checking against the regions excluded by dark photon searches in
LHCb, BaBar, NA48, NA64, E141 and $\nu$-CAL I\@.
Finally, we required our DM candidate, $W^{\prime (p,m)}$, to have the
correct relic density measured by Planck satellite with a spin-independent direct detection cross section
below the limits found by CRESST-III, DarkSide-50 and XENON1T.

In our numerical analysis, we uncovered some interesting features of the
parameter space, such as a correlation between the couplings $g_H$ and $g_X$ 
with superweak size $O(10^{-5} - 10^{-3})$
and that most of our results lie inside the gap between various dark photon
explorations.
The latter becomes more important when we consider that this gap is projected
to be further explored in the future with the upgrades to NA64 and AWAKE and
the B-factory Belle II\@.
This would reduce the allowed parameter space nearly by half, and therefore 
remove a good portion of our allowed parameter space in the lighter $A'$ region.
Moreover, future DM direct detection experiments like NEWS-G and SuperCDMS
may reduce the parameter space by exploring the regions with heavier
$W^{\prime (p,m)}$ and larger cross section.

To summarize, we found that the simplified G2HDM developed in this work
provides a viable vector DM candidate with mass down to
$\mathcal{O}(10^{-2})$~GeV\@.
All the predictions in the scalar and gauge sectors are in good agreement with
current observations.
Importantly, both new vector states, $Z'$ and $A'$, play key roles for DM observables.
Besides the possibility of detecting the $W^{\prime (p,m)}$ in DM direct
detection experiments, the dark photon, $A'$, is predicted to be well
positioned for future observations that may reach $m_{A'} \sim 0.1$~GeV\@.
This work demonstrates that the G2HDM is not only a successful and competitive dark matter
formulation but can also serve as a starting point with diverse exploration possibilities.

\section*{Acknowledgments}

The analysis presented here was done using the resources of the
high-performance T3 Cluster at the Institute of Physics, Academia Sinica.
This work is supported in part by the Ministry of Science and Technology of
Taiwan under Grants No. 108-2112-M-001-046 (TCY) and No. 109-2811-M-001-595 (RR)
and by National Natural Science Foundation of China under Grants No. 11775109 (VQT).
VQT would like to thank the Institute of Physics, 
Academia Sinica, 
Taiwan for its hospitality during this work.

\clearpage

\appendix

\section{\label{appendixA}Mass Spectra of Goldstone Bosons 
and Gauge Fixings in General Renormalizable Gauge}

In Sec.~\ref{sec:scalar_mass} as well as in all previous works, we have derived the scalar and vector boson 
mass spectra in the 't Hooft-Landau  gauge in which all the gauge parameter parameters $\xi$s were set to zero. 
This hides away the issue of gauge dependence and all the Goldstone bosons are massless. Since the physical 
dark Higgs $D$ is a linear combination of $H_2^{0 *}$ in the inert doublet $H_2$ and Goldstone boson $G^p_H$ in the hidden doublet $\Phi_H$, 
one might wonder what would happen in the general renormalizable gauge. We take the opportunity here to examine
this question more careful and discuss the related issue of gauge mixings in G2HDM\@.

\subsection{$G^\pm$ and $G^{(p,m)}_H$}

First $G^\pm$. This is the same as in SM, which after SSB the covariant kinetic term of the Higgs field $H_1$
contains the following mixing term
\be
\label{GWMixing}
- \frac{i}{2} g v W^+_\mu \partial^\mu G^- + {\rm H.c.}
\ee
In the general renormalizable gauge, one introduces the following gauge fixing term
\be
\label{GF:SMW}
-\frac{1}{\xi_W} \Bigl\vert \partial^\mu W_\mu^+ - \frac{i}{2} \xi_W g v G^+ \Bigr\vert^2 \; ,
\ee
where $\xi_W$ is an arbitrary gauge parameter.
Expanding out Eq.~(\ref{GF:SMW}), up to a total derivative, it will cancel the mixing term in Eq.~(\ref{GWMixing}).
In addition, it will modify the $W^\pm$-boson propagator to be $\xi_W$-dependent whose
form is well known in the literature,
and the Goldstone boson $G^\pm$ (absorbed by the longitudinal component of $W^\pm$) 
will develop a mass equals to $gv\sqrt{\xi_W}/2$. 

The story of $G^{(p,m)}_H$ in G2HDM is a little bit more interesting. The mixing term contains two contributions coming from the 
covariant kinetic terms of the $H_2$ and $\Phi_H$ fields
\be
\label{GpmWpmMixing}
-\frac{i}{2} g_H W^p_\mu \left( v_\Phi \partial^\mu G^m_H - v \partial^\mu  H^0_2 \right) \; ,
\ee
which indicates that the physical Goldstone field $\tilde G^{p}_H$ ($\tilde G^{m}_H$) 
is actually a linear combination of $H^{0*}_2$ ($H^{0}_2$ ) and $G^{p}_H$ ($G^{m}_H$), {\it viz}.
\be
\tilde G^p_H = \frac{1}{\sqrt{v^2 + v_\Phi^2}} \left( v_\Phi G^p_H - v H^{0 *}_2 \right)  \; , \;\; 
\tilde G^m_H = \frac{1}{\sqrt{v^2 + v_\Phi^2}} \left( v_\Phi G^m_H - v H^0_2 \right) \; .
\ee
${\tilde G}^{(p,m)}_H$ are absorbed by the longitudinal components of $W^{\prime (p,m)}$. 
The other physical orthogonal combination is the complex dark Higgs $D$
\be
D = \frac{1}{\sqrt{v^2 + v_\Phi^2}} \left( v G^p_H + v_\Phi H^{0 *}_2 \right)  \; , \;\; 
D^* = \frac{1}{\sqrt{v^2 + v_\Phi^2}} \left( v G^m_H + v_\Phi H^0_2 \right) \; .
\ee
In analogous with Eqs.~\eqref{GWMixing} and~\eqref{GF:SMW}, to cancel the mixing term in Eq.~\eqref{GpmWpmMixing}
one introduces the following general gauge fixing term
\be\label{GF:G2HDMWp}
-\frac{1}{\xi_{W^\prime}} \Bigl\vert \partial^\mu W_\mu^p - \frac{i}{2} \xi_{W^\prime} g_H  \tilde v \tilde G^p_H \Bigr\vert^2 \;,
\ee
where $\xi_{W^\prime}$ is an arbitrary gauge parameter and $\tilde v = \sqrt{ v^2  + v^2_\Phi }$. 
As in the case of $W^\pm$, the propagator of $W^{\prime (p,m)}$ is then get modified under this general renormalizable gauge. The 
mass matrix ${\mathcal M}^2_{S'}$ in Eq.~(\ref{eq:goldstonemassmatrix}) is also modified as
\be
{\mathcal M}_{S'}^{ 2} =
\begin{pmatrix}
 \frac{1}{2}\lambda^\prime_{H\Phi}v^2  +  \frac{1}{4} \xi_{W^\prime} g_H^2 v_\Phi^2      &  \frac{1}{2}\lambda^\prime_{H\Phi}v v_\Phi -  \frac{1}{4} \xi_{W^\prime} g_H^2  v v_\Phi \\
 \frac{1}{2}\lambda^\prime_{H\Phi}v v_\Phi -  \frac{1}{4} \xi_{W^\prime} g_H^2 v v_\Phi  &   \frac{1}{2}\lambda^\prime_{H\Phi}v_\Phi^2 
 +  \frac{1}{4} \xi_{W^\prime} g_H^2 v^2 
\end{pmatrix} \; ,
\label{eq:goldstonemassmatrix2}
\ee
which has two eigenvalues $g_H^2 \xi_{W^\prime} \tilde v^2/4$ and $\lambda^\prime_{H\Phi} \tilde v^2/2$. 
The first one is the mass-squared of the Goldstone boson $\tilde G^{(p,m)}_H$ in the general gauge, while the second one is 
the mass-squared of the complex dark Higgs $D$ which is the same as Eq.~(\ref{eq:Dscalarmass}) previously derived 
from the  't Hooft-Landau  gauge as it should!

\subsection{$G^0$, $G^0_H$ and $\mathcal S$}

Similarly for the neutral gauge bosons, after SSB, we have the following mixing terms from the covariant kinetic terms 
of the $H_1$ and $\Phi_H$ fields
\bea
\label{NGBMixings}
&-&  \!\! \frac{1}{2} v \left( g^\prime B^\mu - g W^{3 \mu} + g_H W^{\prime 3 \mu} + 2 g_X X^\mu \right) \partial_\mu G^0 \nonumber \\
&+&  \!\! \frac{1}{2}  v_\Phi \left( g_H W^{\prime 3 \mu} - 2 g_X X^\mu \right) \partial_\mu G^0_H \; .
\eea
For the $U(1)_X$ gauge field, we also use the Stueckelberg mechanism to provide a mass. 
The Stueckelberg Lagrangian is
\bea
\label{Lst}
{\mathcal L}_{\rm St} & = & \frac{1}{2} \left( \partial^\mu \mathcal S + M_X X^\mu \right)^2 \; , \nonumber \\
& = & \frac{1}{2} \left( \partial^\mu \mathcal S \right)^2 + \frac{1}{2} M_X^2 X^\mu X_\mu + M_X X^\mu \partial_\mu \mathcal S \; .
\eea
Here $\mathcal S$ and $M_X$ are the Stueckelberg field and mass respectively. The middle term gives $X$ a mass $M_X$, while 
the last term indicates the mixing of $\mathcal S$ with the longitudinal component of $X^\mu$. Note that $\mathcal S$ is massless.

Recall that the first two terms in Eq.~(\ref{NGBMixings}) are SM-like and they can be combined as
\be
(g^\prime B_\mu - g W^3_{\mu}) = - (g^\prime \sin \theta_W + g \cos \theta_W) Z^{\rm SM}_{ \mu} \; .
\ee 
Thus the total mixing terms including both Eq.~(\ref{NGBMixings}) and the last term in Eq.~(\ref{Lst}) are 
\bea
\label{NGBMixings-All}
&+&  \!\! \frac{1}{2} v \left[ (g^\prime \sin \theta_W + g \cos \theta_W) Z^{{\rm SM} \, \mu}  - g_H W^{\prime 3 \mu} 
- 2 g_X X^\mu \right]\partial_\mu G^0 \nonumber \\
&+&  \!\! \frac{1}{2}  v_\Phi \left( g_H W^{\prime 3 \mu} - 2 g_X X^\mu \right) \partial_\mu G^0_H + M_X X^\mu \partial_\mu \mathcal S \; .
\eea
Note that, as one would expect,
the photon field $A^\mu$ doesn't enter in Eq.~(\ref{NGBMixings-All}) which is coming entirely from SSB and Stueckelberg mechanism. 
The photon field remains massless, it has no associated Goldstone boson and its general gauge fixing term is simply given by 
$-\left( \partial_\mu A^\mu \right)^2 / 2 \xi_\gamma$ as in the SM case. So does the massless gluon field, whose gauge fixing term is 
$-\left( \partial_\mu A^{a \mu} \right)^2 / 2 \xi_g$ where $a=1,\ldots,8$ is the adjoint index of the color group $SU(3)_C$.

As mentioned in the text, the neutral vector gauge bosons $Z^{\rm SM}$, $W^{\prime 3}$ and $X$ are in general mixed together 
according to 
\be
\label{Oij2}
\begin{pmatrix}
Z^{\rm SM} \\
W^{\prime 3}\\ 
X
\end{pmatrix} 
= {\cal O} \cdot 
\begin{pmatrix}
Z_1 \\ Z_2 \\ Z_3 
\end{pmatrix} 
 \; ,
\ee
where $\cal O$ is an orthogonal matrix. 
In our numerical scan of the parameter space in this work, we have 
$Z_1 = Z \simeq Z^{\rm SM}$ which is very close to the SM $Z$-boson,
and $Z_2 = Z^\prime$ and $Z_3 = A^\prime$, both of which are lighter than the $Z$.
We will use $Z_i$ in this appendix instead of $Z$, $Z^\prime$ and $A^\prime$.

In terms of the physical fields $Z_i$, Eq.~(\ref{NGBMixings-All})  becomes
\bea
\label{NGBMixings-All-2}
&+&  \!\! \frac{1}{2} v \left[ (g^\prime \sin \theta_W + g \cos \theta_W) {\mathcal O}_{1i} Z_i^{\mu}  - 
g_H {\mathcal O}_{2i} Z_i^{\mu} - 2 g_X {\mathcal O}_{3i} Z_i^{\mu} \right] \partial_\mu G^0 \nonumber \\
&+&  \!\! \frac{1}{2}  v_\Phi \left( g_H {\mathcal O}_{2i} Z_i^{\mu} - 2 g_X {\mathcal O}_{3i} Z_i^{\mu} \right) \partial_\mu G^0_H 
+ M_X {\mathcal O}_{3i} Z_i^{\mu} \partial_\mu \mathcal S \nonumber \\
& \equiv & 
\sum_{i=1}^3 Z_i^\mu \partial_\mu G_i
 \; .
 \eea
Here we have defined three physical Goldstone fields absorbed by the longitudinal components of the three physical $Z_i$ as
\be
G_i  =  {\widetilde C}_{i1} G^0 + {\widetilde C}_{i2} G^0_H + {\widetilde C}_{i3} \mathcal S \;\;\;\; (i=1,2,3) \; ,
\ee
where the coefficients ${\widetilde C}_{ij}$ can be read off from the first two lines in Eq.~(\ref{NGBMixings-All-2}), namely
\bea
{\widetilde C}_{i1} &= & \frac{1}{2} v \left[ (g^\prime \sin \theta_W + g \cos \theta_W ) {\mathcal O}_{1i} 
- g_H {\mathcal O}_{2i} - 2 g_X {\mathcal O}_{3i} \right] \; , \\
{\widetilde C}_{i2} &= &  \frac{1}{2}  v_\Phi \left( g_H {\mathcal O}_{2i} - g_X  {\mathcal O}_{3i} \right)  \; , \\
{\widetilde C}_{i3} &= & M_X {\mathcal O}_{3i} \; .
\eea

To cancel the mixing term in the last expression in Eq.~(\ref{NGBMixings-All-2}), we need the following gauge fixing term,
\be
- \frac{1}{2 \xi_{i}} \left( \partial_\mu Z^\mu_i - \xi_{i} G_i \right)^2 \; ,
\ee
where $\xi_{i}$ are three arbitrary gauge parameters. 
This gauge fixing term will not only modify the propagator of $Z_i$ but also 
induce a $3 \times 3$ mass mixing matrix ${\mathcal M}^2_{G}$ among the three fields  
$\{ G^0, G_H^0, \mathcal S \}$ to be $\xi_{i}$-dependence, namely
\be
{\mathcal M}^2_{G} = 
\begin{pmatrix}
{\mathcal M}^2_{11} & {\mathcal M}^2_{12} & {\mathcal M}^2_{13} \\
{\mathcal M}^2_{21} & {\mathcal M}^2_{22} & {\mathcal M}^2_{23} \\
{\mathcal M}^2_{31} & {\mathcal M}^2_{32} & {\mathcal M}^2_{33}
\end{pmatrix} \;\;\; ,
\ee
with the following matrix elements
\be
\begin{matrix}
{\mathcal M}^2_{11} = \xi_1 {\widetilde C}_{11}^2 + \xi_2 {\widetilde C}_{21}^2 + \xi_3 {\widetilde C}_{31}^2  \; , \\
{\mathcal M}^2_{12} =  \xi_1 {\widetilde C}_{11} {\widetilde C}_{12} + \xi_2 {\widetilde C}_{21} {\widetilde C}_{22} + \xi_3 {\widetilde C}_{31} {\widetilde C}_{32}  = {\mathcal M}^2_{21} \; , \\
{\mathcal M}^2_{13} = \xi_1 {\widetilde C}_{11} {\widetilde C}_{13} + \xi_2 {\widetilde C}_{21} {\widetilde C}_{23} + \xi_3 {\widetilde C}_{31} {\widetilde C}_{33} = {\mathcal M}^2_{31} \; , \\
{\mathcal M}^2_{22} = \xi_1 {\widetilde C}_{12}^2 + \xi_2 {\widetilde C}_{22}^2 + \xi_3 {\widetilde C}_{32}^2 \; , \\
{\mathcal M}^2_{23} =  \xi_1 {\widetilde C}_{12} {\widetilde C}_{13} + \xi_2 {\widetilde C}_{22} {\widetilde C}_{23} + \xi_3 {\widetilde C}_{32} {\widetilde C}_{33} = {\mathcal M}^2_{32} \; , \\
{\mathcal M}^2_{33} = \xi_1 {\widetilde C}_{13}^2 + \xi_2 {\widetilde C}_{23}^2 + \xi_3 {\widetilde C}_{33}^2
\; .
\end{matrix}
\ee

While the eigenvalues of ${\mathcal M}^2_{G}$ are complicated functions of the gauge fixing parameters, its determinant 
takes a simple form,
\bea
{\rm Det} {\mathcal M}^2_{G} & = & \xi_1 \xi_2 \xi_3 \biggl( 
 {\widetilde C}_{11} {\widetilde C}_{23} {\widetilde C}_{32}
+  {\widetilde C}_{12} {\widetilde C}_{21} {\widetilde C}_{33}
+  {\widetilde C}_{13} {\widetilde C}_{22} {\widetilde C}_{31} \biggr. \nonumber \\
&& \;\; \;\;\; \;\; \;\, \biggl. -   {\widetilde C}_{13} {\widetilde C}_{21} {\widetilde C}_{32}
-  {\widetilde C}_{11} {\widetilde C}_{22} {\widetilde C}_{33}
-  {\widetilde C}_{12} {\widetilde C}_{23} {\widetilde C}_{31}
\biggr)^2 \; .
\eea
Thus as any one of the gauge fixing parameters $\xi_i \to 0$, the corresponding Goldstone boson $G_i$ has vanishing mass, 
reproducing the result of  't Hooft-Landau  gauge.

This completes our discussions of all the gauge fixing terms in G2HDM\@. 
With these gauge fixing terms at hand, one can straightforwardly obtain the 
gauge fixing functions and the procedure of Faddeev-Popov quantization can be proceeded as usual.
Of course physical observables if computed correctly should be independent of all the otherwise arbitrary 
gauge fixing parameters $\xi$s discussed in this Appendix! 
In an ideal world, one would compute things using arbitrary $\xi$s and show the dependence of $\xi$s 
is completely dropped out at the end for any physical observable. In practice, things are hardly get done
that way.


\end{document}